\documentclass[%
 reprint,
 amsmath,amssymb,
 aps,
prl,
nofootinbib,
prstper,
floatfix,
]{revtex4-1}

\usepackage{graphicx}
\usepackage{dcolumn}
\usepackage{bm}
\usepackage{color}
\usepackage[colorlinks]{hyperref}
\usepackage{dsfont}
\usepackage{pst-node}%
\usepackage{array,mathtools,amssymb,booktabs}
\usepackage[utf8]{inputenc}
\usepackage[T1]{fontenc}

    \def \k{\kappa}

\def \k{\kappa}

\def \>{\rangle} 
\def \<{\langle} 
 
\def\be{\begin{equation}} 
\def\ee{\end{equation}} 
\def\longrightharpoonup{\relbar\joinrel\rightharpoonup}
\def\longleftharpoondown{\leftharpoondown\joinrel\relbar}

\def\longrightleftharpoons{
  \mathop{
    \vcenter{
      \hbox{
      \ooalign{
        \raise1pt\hbox{$\longrightharpoonup\joinrel$}\crcr
	  \lower1pt\hbox{$\longleftharpoondown\joinrel$}
	  }
      }
    }
  }
}

\newcommand \bea {\begin{eqnarray}} 
\newcommand \eea {\end{eqnarray}}

\begin{document}

\title[]{Cross-feeding shapes both competition and cooperation in microbial ecosystems}

\author{Pankaj Mehta}%
\email{pankajm@bu.edu}
\affiliation{Department of Physics, Boston University, 590 Commonwealth Avenue, Boston, MA 02139}
\date{\today}

\author{Robert Marsland III}
\email{current address: Pontifical University of the Holy Cross, Rome, Italy}
\affiliation{Department of Physics, Boston University, 590 Commonwealth Avenue, Boston, MA 02139}


\begin{abstract}
Recent work suggests that cross-feeding -- the secretion and consumption of metabolic biproducts by microbes -- is essential for understanding microbial ecology. Yet how cross-feeding and competition combine to give rise to ecosystem-level properties remains poorly understood. To address this question, we analytically analyze the Microbial Consumer Resource Model (MiCRM), a prominent ecological model commonly used to study microbial communities. Our mean-field solution exploits the fact that unlike replicas, the cavity method does not require the existence of a Lyapunov function. We use our solution to derive new species-packing bounds for diverse ecosystems in the presence of cross-feeding, as well as simple expressions for species richness and the abundance of secreted resources as a function of cross-feeding (metabolic leakage) and competition. Our results show how a complex interplay between competition for resources and cooperation resulting from metabolic exchange combine to shape the properties of microbial ecosystems. 
\end{abstract}

\maketitle
\newpage
\clearpage


Microbial communities are found everywhere on the globe from hot springs to human bodies \cite{thompson2017communal, integrative2019integrative}. Large-scale surveys suggest that a these microbial communities are extremely diverse, with hundreds to thousands of 
distinct microbial species all coexisting in a single environment. Recent lab experiments suggest that that diverse communities can also stably coexist even in simple environments with just a handful of externally supplied resources suggesting that classical ecological theories of competition must be modified to describe microbial ecosystems  \cite{goldford2018emergent, enke2019modular, gralka2020trophic, Dal2021Resource, estrela2021nutrient}.  

It is now believed that metabolic interactions
facilitated by cross-feeding -- the secretion and consumption of metabolic biproducts by microbes -- play
a fundamental role in shaping the structure and function of microbial ecosystems. \cite{goldford2018emergent,pacheco2019costless,  amarnath2021stress,Dal2021Resource, pacheco2021non, lechon2021role}.This has led to a substantial body of work trying to extend classical ecological model to incorporate
cross-feeding interactions \cite{goldford2018emergent, butler2018stability, marsland2019available, goyal2018multiple, goyal2021ecology, Dal2021Resource}. One prominent example of this is the Microbial Consumer Resource Model (MiCRM) which extends classical ecological consumer resource models \cite{mac1969species, macarthur1967limiting,tilman1982resource, chesson1990macarthur, chase2009ecological} by including cross-feeding interactions between consumers \cite{ marsland2019available,marsland2020community,fant2021eco}. The MiCRM has been successfully used to explain both laboratory experiments and the natural patterns found in large-scale surveys \cite{marsland2020community, marsland2020minimal, gralka2020trophic, fant2021eco}.

Whereas classical consumer resource models emphasize the importance of competition in shaping ecosystems \cite{tilman1982resource, chesson1990macarthur, chase2009ecological}, the MiCRM includes both competition for resources as well as cooperation between microbes facilitated by cross-feeding interactions. A fundamental question in microbial ecology is to understand how this interplay between competition and cooperation shapes the properties of diverse communities \cite{lechon2021role}. Answering this question has proven difficult due to a lack of analytic results. The MiCRM,  in contrast with many other ecological models, cannot be cast as an optimization problem \cite{mac1969species, macarthur1967limiting, mehta2019constrained, marsland2020minimum} and for this reason is not amenable to analysis using the replica method \cite{tikhonov2017collective, biroli2018marginally, altieri2021properties}.  Here, we use the cavity method, another technique from the statistical physics of disordered systems \cite{advani2018statistical, bunin2017ecological, barbier2018generic}, to derive a mean-field theory for the steady-state solutions for the MiCRM. Unlike replicas, the cavity method does not require the existence of an optimization functional making it ideally suited for analyzing the MiCRM \cite{pearce2020stabilization}. Since this calculation is quite technically involved, we limit ourselves in the main text to presenting the results and implications. A detailed derivation is provided in the appendix.

\begin{figure}[t]
\centering
\includegraphics[width=0.95\columnwidth]{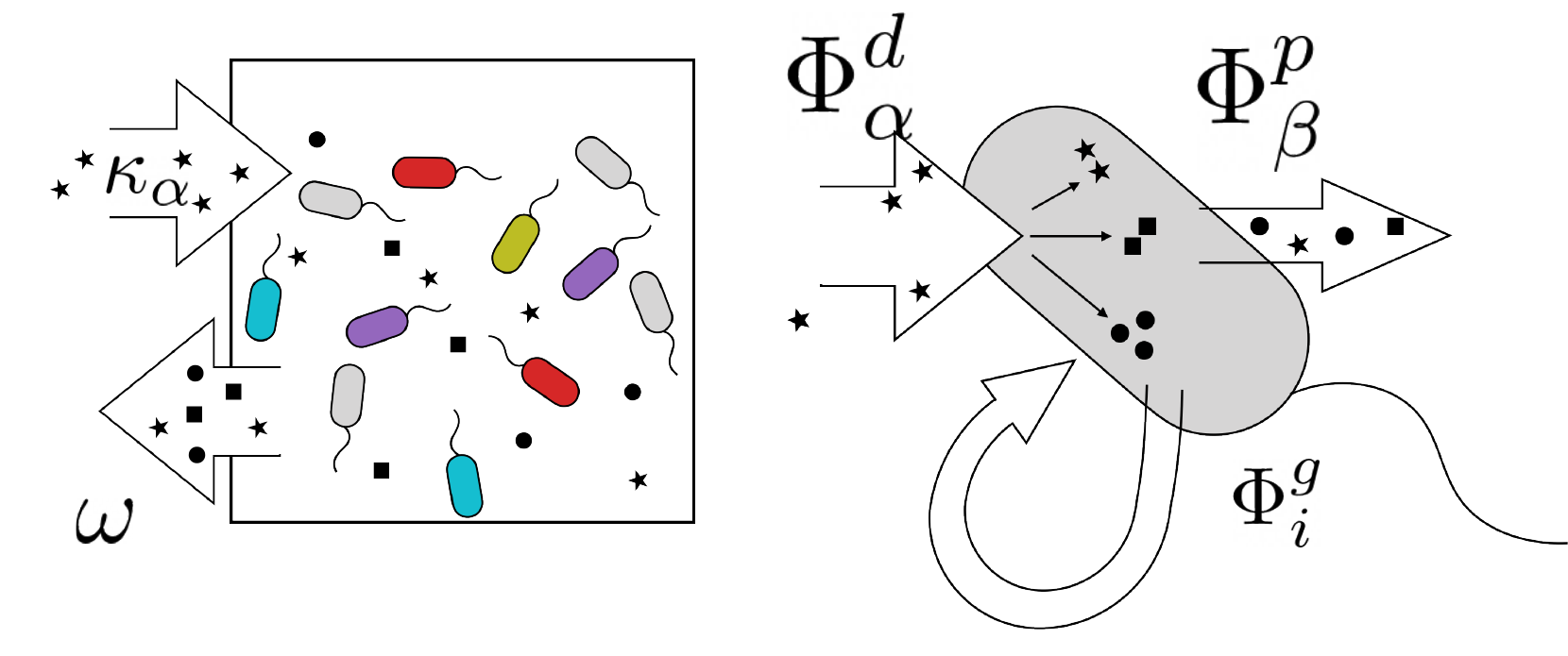}
\caption{{\bf Microbial Consumer Resource Model (MiCRM).} 
Resources $\{R_\alpha\}$ are supplied
externally into the environment at a rate $\kappa_\alpha$ and degraded at a rate
$\omega$.  Microbes (also called consumers) with 
abundances $\{ N_i\}$ consume resources from the environment ($\Phi_\alpha^d$),
using some of the energy to grow ($\Phi_i^g$), while the remaining energy is secreted
back into the environment as metabolic biproducts ($\Phi_\beta^p$). 
We assume there is a single externally supplied resource ($\alpha=E$) so
that $\kappa_\alpha =0$ if $\alpha \neq E$.}
\label{fig:scheme}
\end{figure}

\begin{figure}[t]
\centering
\includegraphics[width=1.0\columnwidth]{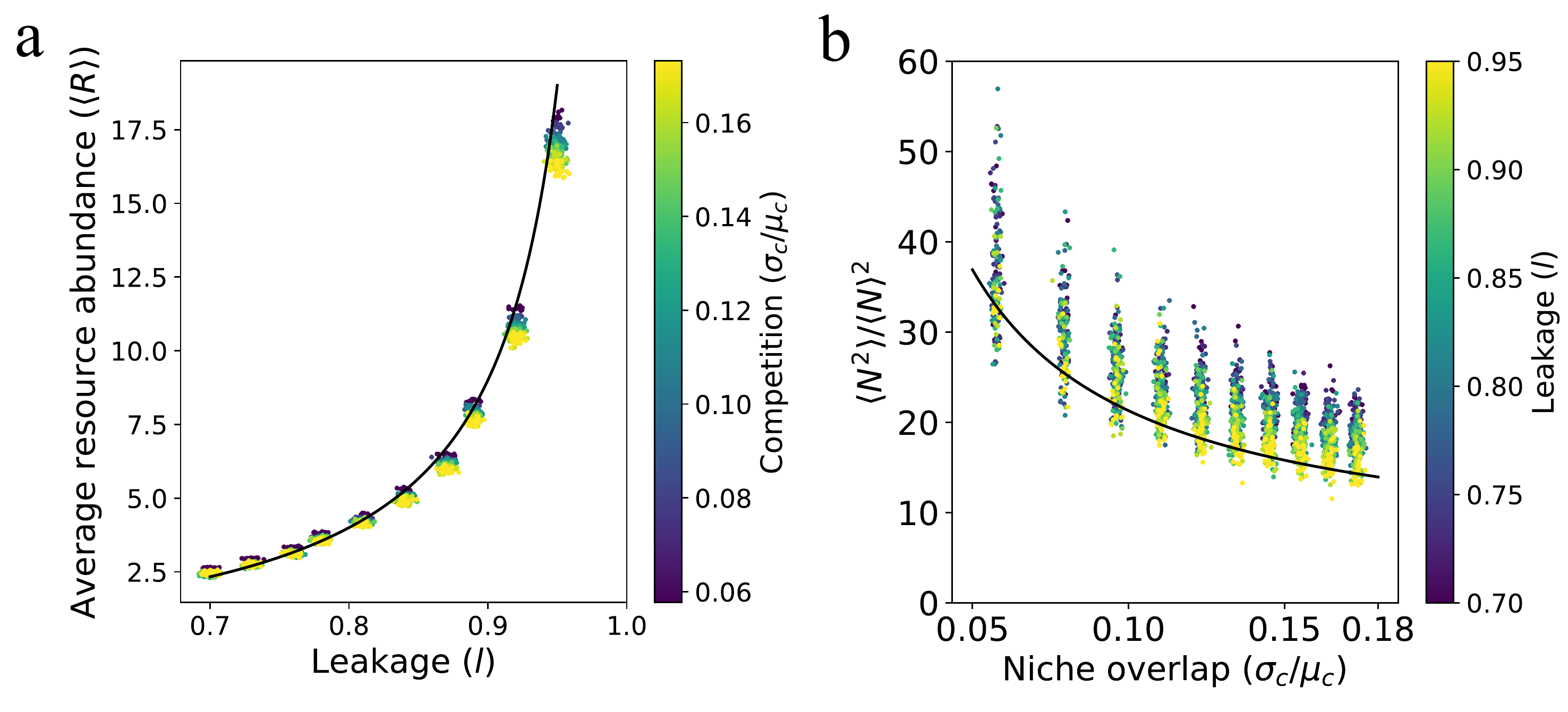}
\caption{{\bf Resources and species abundances are controlled 
by different processes.} (a) Numerical
simulations of average resource abundances as a function of leakage
rate $l$ (x-axis) and $\sigma_c/\mu_c$ (color bar). Solid
line is analytic expression from Eq. \ref{Eq:R_approx_main}. (b)  Numerical
simulations of species $\< N^2 \> /\< N \>^2$ as a function of $\sigma_c/\mu_c$
(x-axis) and leakage rate $l$  (color bar). Solid
line is analytic expression given by Eq. \ref{Eq:Species_approx_main}. 
 (see Fig. S1 for further numerical checks and appendix for details and parameters).}
\label{fig:analytics}
\end{figure}

\noindent {\bf Model.} We briefly summarize Microbial consumer resource model (MiCRM) (see \cite{ marsland2019available,marsland2020community} and appendix for details). Species $i =1 \ldots S$ with abundances $N_i$ can consume $M$ potential resources $R_\beta$ ($\beta=1\ldots M$) that can exist in the environment. A fraction $(1-l)$ of the energy in consumed resources is used for growth while the remaining energy fraction $0 \le l\le 1$ is released back into the environment as metabolic biproducts (see Fig. \ref{fig:scheme}). This is described by the equation:
\bea
{d N_i \over dt} &=N_i \left(\Phi^g_i -m_i \right)
\label{Micro_CRM_N_main}\\
\Phi^g_i &=  \sum_{\alpha} (1-l) w_\alpha g_{i \alpha} c_{i \alpha} R_{\alpha} \nonumber
\eea
 where $w_\alpha$ is the value of one unit of resource to a species (e.g. ATPs that can be extracted); $c_{i \alpha}$ is the rate at which species $i$ consumes resource $\alpha$, $m_i$ is the minimum amount of resources that must be consumed in order to have a positive growth rate, and $g_{i \alpha}$ is a factor that converts consumption into a growth rate for species $i$ when consuming resource $\alpha$. To derive our analytic solution, we consider the case when the consumer preferences $c_{i \alpha}$ are drawn randomly with
$\< c_{i \alpha}\>= {\mu_c \over M}$ and $\<c_{i \alpha} c_{j \beta}\>= {\sigma_c^2 \over M} \delta_{ij} \delta_{\alpha \beta} + {\mu_c^2 \over M^2} \approx {\sigma_c^2 \over M} \delta_{ij} \delta_{\alpha \beta}$. We have numerically checked that in the thermodynamic limit where $M,S \rightarrow \infty$ with $\gamma=M/S$ fixed that our analytic expressions hold even when $c_{i \alpha}$ are drawn from distributions where the $c_{i \alpha}$ are strictly positive (see Figures \ref{fig:distributions} and \ref{fig:analytics}).

In the MiCRM, the resources $R_\alpha$ satisfy their own dynamical equations:
\bea
{d R_{\alpha} \over dt} &=& \k_\alpha - \omega_\alpha R_\alpha - \Phi_\alpha^{d} +\Phi_\alpha^p \\
\Phi_\alpha^d  &=& \sum_j c_{j \alpha} N_j R_\alpha, \hspace{0.1in}
\Phi_\alpha^p = \sum_{j,\beta}  l_\beta {w_\beta \over w_\alpha} D_{\alpha \beta} c_{j \beta}R_\beta N_j. \nonumber
\label{MiCRM_R_main}
\eea
The first two terms on the right hand side of the top equation describe the dynamics of $R_\alpha$ in the absence of microbes, $\Phi_\alpha^d$ describes resource consumption,  and $\Phi_\alpha^p$ the production of resources due to cross feeding. The 
cross-feeding matrix $D_{\alpha \beta}$ encodes the fraction of energy leaked from resource $\beta$ that goes into resource $\alpha$. Motivated by the universality of metabolism, we assume that the $D_{\alpha \beta}$
is the same for all species $i$ and each row can be described using  a Dirichlet distribution. To ensure a good thermodynamic limit, we further assume that elements of $D_{\alpha \beta}$  for a fixed $\beta$ follow a Dirichlet distribution with shape parameters $\{\alpha_j \}$ which scale like $\alpha_j \sim \sqrt{M}$. Under this assumption, to leading order in $M$ we can ignore the anti-correlations between the elements of $D_{\alpha \beta}$  when deriving our mean-field equations and instead approximate the the elements of $D_{\alpha \beta}$ as independent, normal random variables with  $\< D_{\alpha \beta} \> = {1 \over M}$ and $\<D_{\alpha \beta} D_{\gamma \eta} \> =\delta_{\alpha \gamma} \delta_{\beta \eta} {\sigma^2_D \over M}$ (see appendix). We emphasize that our numerical simulations simulate the full dynamics and do not rely on this approximation.

In what follows we set $w_\alpha=w$  and $g_{i\alpha} = g$  for all $\alpha$ and $i$. We also assume that the $m_i$ are independent normal variables with mean $m$ and variance $\sigma_m^2$. Finally, inspired by recent experiments \cite{goldford2018emergent, enke2019modular, gralka2020trophic, Dal2021Resource, estrela2021nutrient} we focus on the case where just one of the resources $R_E$ is supplied externally at a rate $\k_E=\tilde{\kappa}_E M$, and all the other resources result from cross-feeding (i.e $\k_\alpha = 0$ for $\alpha \neq E$). This corresponds to the high-energy regime that
supports diverse communities in \cite{marsland2019available}.


\noindent {\bf Parameters, competition, and cross-feeding.} We are especially interested in understanding
the interplay between competition and cooperation. In all consumer resource models, consumers/microbes that have similar preference for resources compete more than those who have very different consumer preferences since they occupy similar metabolic niches. This intuition can be generalized to diverse communities by looking at the quantity $\sigma_c / \mu_c$ which characterizes the diversity of consumer resource preferences in the regional species pool. When $\sigma_c \ll \mu_c$, all species have the similar consumer preferences and there is lots of competition, whereas when $\sigma_c \gg \mu_c$ there is much less competition between species since species  have distinct consumer preferences. For this reason, we interpret the ratio $\sigma_c/\mu_c$ as a proxy for (the inverse of) the amount of competition in the regional species pool. 

The amount of cross-feeding in the community is controlled by the leakage parameter $l$ which governs the fraction of energy consumed by microbes that is released as metabolic biproducts. When $l=0$, all the energy microbes consume is used for growth and $\Phi_p=0$. In this case, our equations for the MiCRM reduce to the usual equations for a consumer resource model without cross-feeding \cite{tikhonov2017collective, cui2020effect}. In contrast, if $l=1$ all the energy is released as metatabolic biproducts and species cannot grow. More generally, increasing $l$ increases the fraction of energy released as metabolic biproducts and hence serves as a proxy for the amount of cross-feeding in the community \cite{marsland2019available}.


\noindent {\bf Mean-field equations.} In the appendix, we derive mean-field equations for the MiCRM using the cavity method. These calculations are extremely long and technical and here we confine ourselves to simply stating
the final results.  As in the consumer resource model with linear resource dynamics \cite{cui2020effect}, to derive our expressions we assume replica symmetry and solve the cavity-equations using a Taylor expansion that approximates the fluxes $\Phi_i^g$,
$\Phi_\alpha^d$, and $\Phi_\alpha^p$ using their means and variances. We note that the the expectation value $\<X\>$ denotes averages over both random realizations of the parameters (quenched averages) as well as species/resource abundance distributions. The end result of the cavity procedure is a set of self-consistency equations relating the fraction of surviving species $\phi_N$, the mean $\<N\>$ and second-moment $\<N^2\>$ of the species abundance distribution, the mean $\<R\>$ and second moment $\<R^2\>$ of the resources secreted through cross-feeding, and the cavity susceptibilities $\chi$ and $\nu$ (see Appendix for full definitions). 

These equations can be most succinctly written in terms of the average effective degradation rate of resources 
in the environment $\< \omega^{\rm eff}\>=\omega +\mu_c \<N\>$, which is the sum of the dilution rate $\omega$ and the average rate at which microbes consume resources $\mu_c \<N\>$. In addition, we define
the parameters
\begin{align}
q &=l\left( 1-{\omega \over \<\omega_{eff}\>} \right)  \\
\epsilon_d^2 &= \gamma {\sigma_c^2 \over \mu_c^2} {\<N^2\> \over \<N\>^2}, \hspace{0.1in} \epsilon_p^2 = {\sigma_D^2 \over D^2} q^2, \hspace{0.1in} \epsilon_k^2 = \gamma^{-1} \phi_N + {1 \over 2} \epsilon_d^2\nonumber
\label{eq:summary1_main}
\end{align}
which characterize fluctuations in the fluxes appearing in Eq.~\ref{MiCRM_R_main}. In terms of these quantities, the species
distributions satisfy the self-consistency equations
\begin{align}
\phi_N &= w_0(\Delta_N) \\
\langle N\rangle &= \frac{\sigma_{N}}{(1-l) g\sigma_c^2  \<\chi R\> } w_1(\Delta_N)\nonumber \\
\langle N^2 \rangle &= \left( \frac{\sigma_{N}}{(1-l)g\sigma_c^2 \<\chi R\> }\right)^2 w_2(\Delta_N) \nonumber \\
w_n(\Delta)&= \int_{-\Delta}^\infty {1 \over \sqrt{2\pi} }e^{-z^2 \over 2}(z+\Delta)^n \nonumber \\
\Delta_N &= \frac{(1-l)\mu_c g\<R\>-m}{\sigma_{N}} \nonumber \\
\sigma_N^2 &=\sigma_m^2+ (1-l)^2\sigma_c^2g^2 \<R\>^2(1+\epsilon_d^2 + \epsilon_p^2), \nonumber
\label{eq:summary5_main}
\end{align}
The full species abundace distribution can be calculated by noting that probability of observing a species
with abundance $N_0$ in the microbial community is given by a truncated Gaussian described of the form
\be
N_0 = \mathrm{max}\,  \left[0, {(1-l)\mu_c g\<R\>-m + \sigma_N  z_0^g \over (1-l) g \sigma_c^2\<\chi R\> } \right],
\label{Eq:defN0_main}
\ee
where $z_g$ is a Gaussian random normal variable describing fluctuations in the growth flux $\Phi_0^g$ and the equation for the expectation $\<\chi R\>$ is given below.

We also find that the moments of the resource abundances satisfy the self-consistency equations
\begin{align}
\< R\> 
 &=  {\tilde{k}_E \over \<\omega^{eff}\>}{q \over 1-q}\left( 1+ {\epsilon_d^2 \over 1-q} - {(1+ \gamma)\epsilon_k^2 \over 1-q}\right) + \mathcal{O}(\epsilon^3) \nonumber \\
\langle R^2 \rangle &=\<R\>^2( 1+ \epsilon_d^2 + \epsilon_p^2)  +\mathcal{O}(\epsilon^3),
\label{eq:summary3_main}
\end{align},
where $\mathcal{O}(\epsilon^3)$ indicates that we have only kept terms to second order in $\epsilon_d$, $\epsilon_p$, and $\epsilon_k$.  The full resource distribution is given by
\be
\begin{split}
R_\alpha &= \frac{\<R\>}{2\gamma^{-1} \phi_N + \epsilon_d^2} \times\\
&\left[\sqrt{(1+\epsilon_d z_\alpha^d)^2 +2(2\gamma^{-1}\phi_N+\epsilon_d^2)(1+\epsilon_p z_\alpha^p)}
-1-\epsilon_d z_\alpha^d\right],
\end{split}
\label{eq:Rdist_main}
\ee
where $z_\alpha^d$ and $z_\alpha^p$ are standard random normal variables with mean zero
and variance $1$ describing the fluctuations in the depletion and production fluxes.

Finally, these equations must be supplemented by equations for expectations involving the cavity susceptibilities of the form
\begin{align}
\tilde{\nu} \< R\> &={m \over g\mu_c}\left(\gamma^{-1}\phi_N + \frac{1}{2}\epsilon_d^2\right) +\mathcal{O}(\epsilon^3) \nonumber \\
\langle \chi R\rangle &= \frac{\gamma^{-1}\phi_N}{\tilde{\nu}} = \frac{\gamma^{-1}\phi_N\<R\>}{ \< \omega^{\rm eff}\>\left(\gamma^{-1}\phi_N + \frac{1}{2}\epsilon_d^2\right)}.
\end{align}

To check our expressions, we performed numerical simulations of the MiCRM using the Community Simulator package (see appendix and associated Github repository) \cite{marsland2020community}. In these numerical simulations a single externally supplied resource is supplied to the environment, but microbes can produce M=300 additional resources through cross-feeding (leakage rate $l=0.8$). The consumer preference matrix $c_{i \alpha}$ were chose from a binomial distribution. Figure \ref{fig:distributions} shows that the analytic expressions given by Eqs. \ref{Eq:defN0_main} and \ref{eq:Rdist_main} for the species and resource distributions (solid lines) agree well with distributions obtained numerically suggesting the cavity equations are capturing the essential ecology of the MiCRM.

\noindent {\bf Average metabolite abundance depends is controlled by leakage.} One striking result of our analytic expressions is that to leading order in $\epsilon=\{\epsilon_d, \epsilon_p, \epsilon_k\}$ the average abundance of metabolic biproducts $\<R\>$ depends only on the leakage rate $l$ and the effective degradation rate $\<\omega_{eff}\>$ but not on the amount of competition in the regional species pool as measured by $\sigma_c/\mu_c$ (see Eq. \ref{eq:summary3_main}). When $\<\omega_{eff}\> \gg \<\omega\>$, we can use energy conservation to derive a particularly simple approximate expression for the average metabolite abundances
\be
\<R\> \approx  {\<N\>m l/ (1-l) (\<\omega_{eff}\> -\omega)}.
\label{Eq:R_approx_main}
\ee
A comparison of this expression with numerics is shown in Fig. \ref{fig:analytics}a. From this, we conclude that metabolite concentrations are affected by the regional species pool only through the average species abundance $\<N\>$ and average resource degradation rate $\<\omega_{eff}\>$. This shows the resource abundances reflect the functional structure of the community rather than taxonomy (i.e. exactly what microbes are present).  

\noindent {\bf Species distribution shaped by competition in regional species pool.}   The cavity solution also shows that the species abundance distribution is shaped primarily by competition ($\sigma_c/\mu_c$) and depends much more weakly on the amount of cross-feeding as measured by the leakage rate $l$. As shown in the appendix, we find that to leading order in $\epsilon=\{\epsilon_d, \epsilon_p, \epsilon_k\}$, that the quantity $\Delta_N$ depends only on
${\sigma_c^2 /\mu_c}$. Since $\Delta_N$ is the primary determinant of species abundances, our analysis implies that both the fraction of surviving species $\phi_N$ as well as the ratio 
\be
\<N^2\>/\<N\>^2=w_2(\Delta_N)/[w_1(\Delta_N)]^2
\label{Eq:Species_approx_main}
\ee
should depend strongly on the amount of competition $\sigma_c/\mu_c$ but be largely agnostic to $l$. Fig.\ref{fig:analytics}b shows numerical simulations confirming that this is indeed the case. 

\begin{figure}[t]
\centering
\includegraphics[width=0.8\columnwidth]{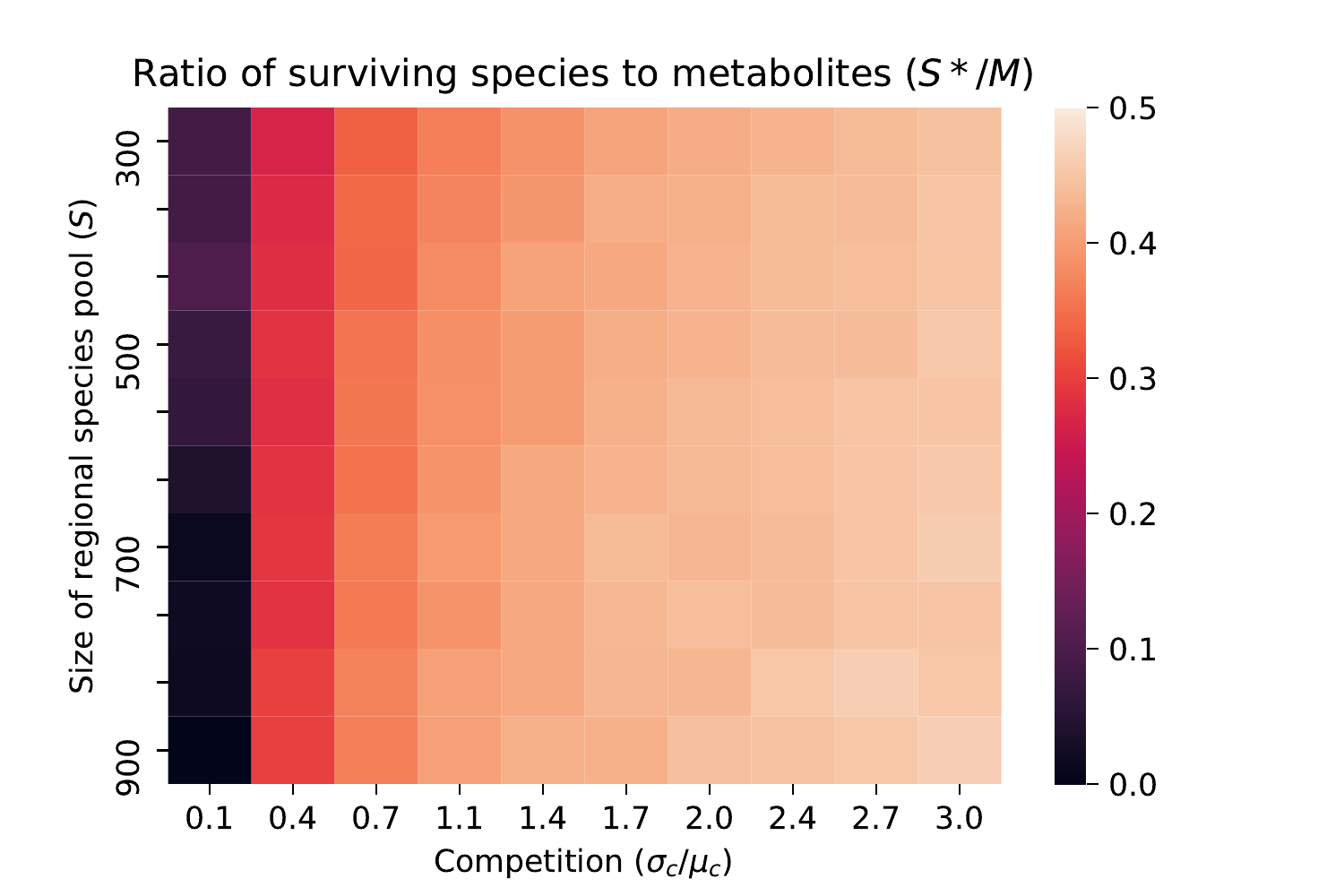}
\caption{{\bf Species packing bound.} Ratio of surviving species 
to resources (including metabolic biproducts) ${S^*/M}$ for different sizes
of regional species pools $S$ and different choices of $\sigma_c/\mu_c$.
Notice that $S^*/M \le 1/2$, consistent analytic bound derived in appendix. Further
numeric checks in Fig. S2. Parameters in appendix.}
\label{fig:bound}
\end{figure}

\noindent {\bf Species packing bound.} One final consequence of our analytic solution is that it allows us to derive a species packing bound for microbial ecosystems \cite{mac1969species, macarthur1967limiting}. In particular, we can ask how many species can co-exist in a microbial ecosystem with cross-feeding? We show in the appendix using a very similar argument to \cite{cui2020effect} that the ratio of surviving species $S^*$ to metabolic biproducts $M$ must be less than half (i.e $S^*/M \le 1/2$) in the thermodynamic limit. Fig. \ref{fig:bound} show numerical simulations confirming this result. These results indicate that the species packing bound found for linear resource dynamics without cross-feeding in \cite{cui2020effect} also holds for the MiCRM where resources are primarily produced by cross-feeding.


\noindent{\bf Discussion.} In this paper, we have used the cavity method to analytically calculate self-consistent mean-field equations for the MiCRM.  Our solution illustrates that an intricate interplay between competition for resources and cooperation resulting from metabolic exchange shape the properties of microbial ecosystems. We found that metabolite abundances are primarily controlled by the leakage  and community-level consumption rates of resources. This finding is consistent with recent experimental and theoretical work suggesting that community-level functional structure is much more conserved in similar environments than taxonomy \cite{louca2016high, louca2016decoupling}. Indeed, our analytic solutions suggest that the metabolite abundances are largely agnostic to which microbes are producing or consuming resources as long as the community is sufficiently diverse. 

In contrast, we find that the species abundance distribution depends primarily on the amount of competition in the regional species pool and is largely agnostic to whether metabolites were supplied externally or internally generated by the community through cross-feeding. This suggests that competition is the primary force shaping taxonomic structure. However, we emphasize that cross-feeding interactions modify the environment by creating new metabolic niches and thus fundamentally shape the terms on which microbes comepete. This basic picture is given further support by our calculation showing that species packing in the MiCRM is analogous to the bound derived for ecosystems without cross-feeding where all resources are supplied externally: the number of surviving species is always less than equal to the number of metabolites in the ecosystem. In other words, species packing in diverse communities is agnostic to how resources are generated.

There are number of promising directions worth pursuing further. In our calculations, both the consumer preferences and cross-feeding matrix were unstructured and we focused on steady-states. It should be possible to extend the calculations presented here to include taxonomic and metabolic structure in the consumption and cross-feeding matrices, as well as metabolic constraints \cite{posfai2017metabolic, caetano2021evolution, pacciani2020dynamic}. More ambitiously, if will be interesting to extend our calculation to derive dynamical mean-field equations \cite{pearce2020stabilization, roy2019numerical}. Another promising direction is to understand how space and migration modify the intuitions found here \cite{weiner2019spatial, gupta2021effective, dukovski2020computation}. Finally, we would like to explore if we can exploit the mean-field equations derived here to understand eco-evolutionary dynamics in the presence of cross-feeding \cite{good2018adaptation}. Recent simulations and analytic arguments suggest that such eco-evolutionary dynamics may depend largely on the functional structure of the community, opening a potential avenue of extending these calculations to include evolutionary dynamics \cite{good2018adaptation,moran2021improve, tikhonov2020model, fant2021eco}.

\noindent {\bf Acknowledgments:} We  thank Wenping Cui, Jason Rocks, and Josh Goldford for useful discussions. This work was supported by NIH NIGMS 1R35GM119461 and a Simons Investigator in the Mathematical
Modeling of Living Systems (MMLS) award to PM. We also acknowledge support from the SSC computing cluster at BU.
P.M. and R.M. contributed equally to this work.
\bibliography{ref.bib}

\pagebreak
\widetext
\appendix
\begin{center}
\textbf{\large Supplemental Appendix}
\end{center}

\appendix
\onecolumngrid
\renewcommand{\thefigure}{S\arabic{figure}}
\setcounter{figure}{0}    

\section{Basic Setup}
We start by briefly summarize Microbial consumer resource model (MiCRM) introduced and developed in \cite{goldford2018emergent, marsland2019available, marsland2020minimal, marsland2020community}. Species, indexed $i =1 \ldots S$, grow by consuming $M$  potential resources, $R_\beta$ ($\beta=1\ldots M$), that can be found in the the environment. In the MiCRM, the consumption of resources is never one-hundred percent efficient and a fraction $l_\alpha$ of the energy in the consumed resources ``leak'' back into the environment as metabolic biproducts.

In this MiCRM, this is described by the equation:
\bea
{d N_i \over dt} &=& N_i \left(\Phi^g_i -m_i \right)
 \nonumber \\
\Phi^g_i &=&  \sum_{\alpha} (1-l_\alpha) \mathrm{w}_\alpha g_{i \alpha} c_{i \alpha} R_{\alpha}
\label{CRM_N}
\eea
 where $\mathrm{w}_\alpha$ is the value of one unit of resource to a species (e.g. ATPs that can be extracted); $c_{i \alpha}$ is the rate at which species $i$ consumes resource $\alpha$, $m_i$ is the minimum amount of resources that must be consumed in order to have a positive growth rate. Finally, $g_{i \alpha}$ is a factor that converts consumption into a growth rate for species $i$ when consuming resource $\alpha$.
 
 In the MiCRM, resources satisfy their own dynamical equations
\bea
{d R_{\alpha} \over dt} &=& \k_\alpha - \omega_\alpha R_\alpha - \Phi_\alpha^{d} +\Phi_\alpha^p \nonumber \\
\Phi_\alpha^d  &=& \sum_j c_{j \alpha} N_j R_\alpha \nonumber \\
\Phi_\alpha^p &=& \sum_{j,\beta}  l_\beta {\mathrm{w}_\beta \over \mathrm{w}_\alpha} D_{\alpha \beta} c_{j \beta}R_\beta N_j
\label{CRM_R}
\eea
where the first two terms describe resource dynamics in the absence of microbes, $\Phi_\alpha^{d}$ describes the depletion of resource $\alpha$ due to consumption, and the production of resources due to cross feeding from metabolic biproducts is encoded in $\Phi_\alpha^p$. 
The matrix $D_{\alpha \beta}$ appearing in $\Phi_\alpha^p$ encodes the fraction of the energy released in metabolic biproduct $\alpha$ when a microbe consumes a unit of resource $\beta$ (see below for more details). We note that these equations reduce to the equations for the original Consumer Resource Model of Levins and MacArthur in the absence of leakage (i.e. $l_\alpha=0$)\cite{mac1969species, macarthur1967limiting,tilman1982resource} .

To proceed, it is helpful to introduce some additional notation and make some simplifications:

\begin{itemize}
\item Define the ratio of potential resources to species by 
\be
\gamma=M/S
\label{defgamma}
\ee
\item For simplicity assume that both the growth rates and resource weights can be assumed to be constant $g_{i\alpha}=g$ and $\mathrm{w}_\alpha=\mathrm{w}=1$.  This is equivalent to the physical assumption that \emph{ all metabolic pathways are basically equally efficient}.
\item We will also assume that all resources are leaked at the same rate so that $l_\alpha=l$ and that the resource dilution rates are all equal with $\omega_\alpha=\omega$.
\item We focus on the case where just one resource $R_E$ is supplied externally, with influx $\k_E$, and all the others are internally generated through biproduct secretion. In other words, $\k_\alpha = 0$ for $\alpha \neq E$. 
\item Finally, all our calculations will be performed in the thermodynamic limit with $M, S \rightarrow \infty$ and the ratio $\gamma$ fixed. For this reason, we will only keep terms that are order $1$ in both $M$ and $S$.
\item We assume that there is no family structure in either the species consumer preferences or the cross-feeding matrix. Even this simple case has been shown to describe many real experiments on microbial communities \cite{marsland2020minimal}.
\end{itemize}

We will consider the case when the consumer preferences $c_{i \alpha}$ are drawn randomly with
\be
\< c_{i \alpha}\>= {\mu_c \over M}
\ee
and
\be
\<c_{i \alpha} c_{j \beta}\>= {\sigma_c^2 \over M} \delta_{ij} \delta_{\alpha \beta} + {\mu_c^2 \over M^2} \approx {\sigma_c^2 \over M}\delta_{ij} \delta_{\alpha \beta},
\ee
where in writing the second line we have ignored terms that are higher order in $M$. Since there are no correlations between the consumer preferences for different species, we have made the ecological assumption that  \emph{there is no family structure} in the regional species pool.

We also assume no mebaolic structure in the cross feeding matrix $D_{\alpha \beta}$. In general, this matrix encodes the fraction of energy leaked from resource $\beta$ that goes into resource $\alpha$. Following earlier works, we assume that each row in the matrix follows a  Dirichlet distribution. To ensure a good thermodynamic limit, we assume that the shape parameters $\{\alpha_j \}$ of the  Dirichlet distribution scale like $\alpha_j \sim \sqrt{M}$ and that the sum of the shape parameters $\alpha_0 = \sum_{j=1}^M \alpha_j$ scales like $M^{3/2}$. With these assumptions, the expectation, variance, and covariance of a variables $\{X_i\}$ draw from a  Dirichlet distribution scale like 
\bea
E[X_i] &=& {\alpha_i \over \alpha_0} \sim {1 \over M} \\
Var[X_i] &=&  {\alpha_i (\alpha_0 -\alpha_i) \over \alpha_0^2 (\alpha_0+1)}  \sim {1 \over M}\\
Cov[X_i, X_j] &=& {-\alpha_i \alpha_j \over \alpha_0^2 (\alpha_0+1)} \sim {1 \over M^2} \approx 0.
\eea
Thus, to leading order in $M$ we can ignore the anti-correlations between elements of the Dirichlet distribution in the thermodynamic limit.
We can translate these observation to the cross-feeding matrix $D_{\alpha \beta}$ to get
\be
\< D_{\alpha \beta} \> = {D \over M} ={1 \over M}
\ee
and
\be
\<D_{\alpha \beta} D_{\gamma \eta} \> =\delta_{\alpha \gamma} \delta_{\beta \eta} {\sigma^2_D \over M}.
\ee
Importantly, we can ignore correlations between the elements of $D_{\alpha \beta}$. Note also that by assumption $D = 1$, since $D_{\alpha\beta}$ is defined as a matrix that partitions a given total quantity of output flux over possible metabolites, with $\sum_\alpha D_{\alpha\beta} = 1$.

\section{Energy Conservation}

One important feature of the MiCRM is that it explicitly accounts for energy conservation. The total power (energy per unit time) that flows into the system is just 
\be
P_{input}= w \kappa_E,
\ee 
where $\kappa_E$ is the rate at which units of the external resource is added to the system and $w$ is the energy contained in each unit of resource. At steady-state, this energy is either consumed by the microbes at a rate $P_{microbe}$ or lost to the environment $P_{environment}$ due to resources disappearing at a rate $\omega_\alpha$ (i.e. we are considering an open system where resource $\alpha$ is lost at a rate $\omega_\alpha$). We know that by construction
\be
P_{microbe}= \sum_i  N_i (1-l_\alpha) \mathrm{w}  c_{i \alpha} R_{\alpha}=\sum_i g^{-1} N_i m_i,
\ee
where in going to the second line we have Eq.~\ref{CRM_N}, noting that at steady-state ${d N_i \over dt}=0$ for all microbes. We also have by definition that
\be
P_{environment}=\sum_{\alpha \neq E} \mathrm{w} \omega_\alpha R_\alpha + \mathrm{w} \omega_E R_E,
\ee
where for future convenience we have explicitly separated the externally supplied resource.
Energy conservation is the statement that
\be
P_{input}=P_{microbe}+P_{environment}
\ee
Plugging in the expression above yields the relationship
\begin{align}
\kappa_E = \omega_E R_E + \sum_{\alpha \neq E} \omega_\alpha  R_\alpha + \sum_i \frac{N_i m_i}{\mathrm{w} g}.
\label{eq:energy1}
\end{align}
In the limit, where $P_{microbe} \gg P_{environment}$ this becomes
\be
\kappa_E  \approx  \sum_i \frac{N_i m_i}{\mathrm{w} g}.
\label{Eq:EnergyCons}
\ee

\section{Externally Supplied Resource}

In our set-up, there is a single distinguished resource $R_E$ that is supplied externally. The dynamics of this resource are given by Eq.~\ref{CRM_R} with $\kappa_E \neq 0$. In fact, to ensure a good thermodynamic limit we must scale $\kappa_E$ with the number of resources $M$ \cite{marsland2019available}. In other words, we have
\be
\kappa_E =\tilde{\k}_E M.
\ee
With this choice, we see that the steady-state abundance of the externally supplied resource will also scale extensively $R_E \sim M$. For this reason, we can solve for the steady-state abundance of $R_E$ using ``naive mean-field theory''. In particular, within naive MFT, the production and depletion flux in Eq.~\ref{CRM_R} can be replaced by their mean-field averages
\bea
\<\Phi_E^d\>  &=&\< \sum_j c_{j E} N_j R_\alpha \> \approx  \gamma \mu_c \<N\> R_\alpha \nonumber \\
\<\Phi_E^p \> &=& \< \sum_{j,\beta}  l D_{E \beta} c_{j \beta}R_\beta N_j \> \approx  \gamma^{-1} l D \mu_c \<N\>\<R\>,
\eea
where we have defined the averages
\bea
\<N \> = {1 \over S} \sum_{j=1}^S N_j \nonumber \\
\<R\> = {1 \over M} \sum_{\alpha=1}^M R_\alpha.
\eea
Plugging these equations into Eq.~\ref{CRM_R} and solving for the steady-state gives
\be
R_E = {\kappa_E + D \mu_c \<N\>\<R\> \over  \omega + \gamma^{-1} \mu_c \<N\>} \approx M {\tilde{\kappa}_E \over  \<\omega^\mathrm{eff}\>},
\label{Eq:REdef}
\ee
where in going to the second line we have dropped sub-leading order term in $M$ and defined the average effective degradation
rate
\be
 \<\omega^\mathrm{eff}\>= \omega + \gamma^{-1} \mu_c \<N\>.
\ee
Thus, as expected the steady-state abundance of the externally supplied resource $R_E$ is extensive in $M$.

\section{Means, variances, and covariances for fluxes of self-generated resources}

In the thermodynamic limit, we treat the production flux $\Phi_\alpha^p$ and depletion flux $\Phi_\alpha^d$ in Eq.~\ref{CRM_R}  for self-generated resources as random variables. Under our assumption of  \emph{replica-symmetry}, such random variables are fully characterized by their means, variances, and covariances. To calculate these quantities, we make use of the cavity method.  In particular, we will consider a system in the \emph{absence} of resource $\alpha$. We will denote the steady-state
abundances reached by species $N_j$ and resources $R_\beta$ in the absence of resource $\alpha$ by $N_{j/ \alpha}$ and $R_{\beta / \alpha}$. Since through out this paper we will only be considering steady-states, our notation makes no distinction between steady-state quantities and dynamical quantities. We will then relate the means and variances of the fluxes $\Phi_\alpha^p$ and $\Phi_\alpha^d$ to $N_{j/ \alpha}$ and $R_{\beta / \alpha}$ using perturbation theory.
The logic behind this somewhat involved cavity method is that unlike naive-MFT this method properly accounts for all correlations in the problem. 
\subsection{Depletion flux}

Let us first write
\be
c_{i \alpha} = {\mu_c \over M} + { \sigma_c \over \sqrt{M} } x_{i \alpha}
\ee
where the $x_{i \alpha}$ are uncorrelated, standard normal variables.  We first find
\be
\Phi_\alpha^d = R_\alpha \gamma^{-1} \mu_c \< N\> +   R_\alpha { \sigma_c \over \sqrt{M} } \sum_j x_{j \alpha}N_{j/\alpha}
\ee
where through an abuse of notation we have defined the expectation value of $\< N \>$ as 
\be
\< N\> = {1 \over S} \sum_j N_{j/\alpha}
\ee
and $N_{j/\alpha}$ is the population of species $j$ in the equilibrium state with resource $\alpha$ absent. This gives us that the mean of the depletion flux is just
\be
\<\Phi_\alpha^d\> = \gamma^{-1}\mu_c \< N\> R_\alpha 
\label{eq:phiDavg}
\ee
We can define
\be
\delta \Phi_\alpha^d= \Phi_\alpha^d-\<\Phi_\alpha^d\> =R_\alpha { \sigma_c \over \sqrt{M} } \sum_j x_{j \alpha}N_{j/\alpha}
\ee

We can also calculate the variance of this flux 
\bea
\sigma_{d, \alpha}^2 \equiv \tilde{\sigma}_{d \alpha}^2 R_\alpha^2 \equiv \< (\delta \Phi_\alpha^d)^2  \> &=&  R_\alpha^2 { \sigma_c^2 \over M } \sum_{j, k}  \<x_{j \alpha}x_{k \alpha} \>N_{j/\alpha} N_{k/\alpha} \nonumber \\
&=&  \gamma^{-1}   \sigma_c^2 \<N^2 \>R_\alpha^2
\label{eq:phiDvar}
\eea
where we have defined 
\be
\< N^2 \> = {1 \over S} \sum_j N_{j/\alpha}^2
\ee
and have defined our shorthand symbol $\tilde{\sigma}_d^2$ for this variance in a way that keeps the $R_\alpha^2$ visible, since it will be important later on.

This allows us to approximate the production flux as a Gaussian random variable with
\be
\Phi_\alpha^d= \<\Phi_\alpha^d\>+ \tilde{\sigma}_{d, \alpha} R_\alpha z_\alpha^d,
\ee
where $z_\alpha^d$ is a unit normal random variable and $ \<\Phi_\alpha^d\>$ and $\tilde{\sigma}_{d, \alpha}$ are defined in Eq.~\ref{eq:phiDavg} and Eq.~\ref{eq:phiDvar} respectively.

\subsection{Production flux}

To calculate the mean and variance of the production flux, we once again start by making the dependence of $D_{\alpha \beta}$ of the $M$ explicit 
by writing
\be
D_{\alpha \beta}= {D \over M} +  {\sigma_D \over \sqrt{M} } y_{\alpha\beta},
\ee
where the $y_{ \alpha\beta}$ are uncorrelated, standard normal variables. This lack of correlation follows from the discussion of the Dirichlet distribution
above. In terms of these variables we have
\bea
\Phi_\alpha^p &=& l \sum_{j, \beta \neq E,\alpha} \left({D \over M} +  {\sigma_D \over \sqrt{M} } y_{\alpha\beta}\right)
\left({\mu_c \over M} + { \sigma_c \over \sqrt{M} } x_{j \beta}\right)R_{\beta/\alpha} N_{j/\alpha} \nonumber\\
&&+
\gamma^{-1} l D\mu_c \<N\>{R_E \over M}
\eea
where once again we have treated the external resource in naive-MFT since it is extensive in $M$.

We have to take care in computing the mean flux, because there is an important correlations among $N_j$, $x_{j\beta}$ and $R_\beta$:
\begin{align}
\<\Phi_\alpha^p\> &=\gamma^{-1} lD \left( \mu_c \<N\>(\<R\>+ {R_E \over M})+\sqrt{M}\sigma_c \<NxR\> \right)
\label{eq:phiPavg}
 \end{align}
where in the last line we have used that fact this second term scales like $\sqrt{M}^{-1/2}$ (see below for self-consistency proof) 
and we have defined
\be
\<R\> = {1 \over M} \sum_\beta R_{\beta/\alpha}
\ee
and
\be
\<NxR\> = {1 \over MS} \sum_{\beta j} N_{j/\alpha} x_{j\beta} R_{\beta/\alpha}. 
\ee
In writing this we have used the fact that $y_{\alpha \beta}$ is not correlated with $R_{\beta/\alpha}$ and $N_{j \alpha}$ and hence all averages involving this quantity can be set to zero. 

In addition to the mean value, to apply the cavity method we will also have to think about the fluctuations around the mean:
\bea
\delta \Phi_\alpha^p &\equiv & \Phi_\alpha^p- \<\Phi_\alpha^p\> \nonumber \\
 &=& \frac{l\sigma_D}{\sqrt{M}}\sum_{j,\beta\neq \alpha} y_{\alpha\beta}\left({\mu_c \over M} + { \sigma_c \over \sqrt{M} } x_{j \beta}\right)R_{\beta/\alpha} N_{j/\alpha} -\gamma^{-1} lD\sqrt{M}\sigma_c \<NxR\> 
 \label{def:deltaPhi_p}
\eea
A straightforward but slightly tedious calculation shows that 
\begin{multline}
\sigma_{p, \alpha}^2 \equiv \<(\delta \Phi_\alpha^p)^2\>= \gamma^{-2} l^2 \sigma_D^2 \left[\mu_c^2 \<R^2\>\<N\>^2+ \sigma_c^2 M(\<N x R^2 x N\> +\<NxR\>^2)+ 2\mu_c \sigma_c \<N\>\sqrt{M}\<x N R^2\> \right].
\label{eq:phiPvar}
\end{multline}
where again in the approximation we have ignored terms that scale as $\sqrt{M}^{-1/2}$ and defined
\bea
\<NxR^2xN\> &=& {1 \over MS^2} \sum_{\beta jk} N_{j/\alpha} x_{j\beta} R_{\beta/\alpha}^2 x_{k\beta}N_{k/\alpha}\\
\<NxR^2\> &=& {1 \over MS} \sum_{\beta j} N_{j/\alpha} x_{j\beta} R_{\beta/\alpha}^2.
\eea

This allows us to approximate the production flux as a Gaussian random variable with
\be
\Phi_\alpha^p= \<\Phi_\alpha^p\>+ \sigma_{p, \alpha} z_\alpha^p,
\ee
where $z_\alpha^p$ is a unit normal random variable and $ \<\Phi_\alpha^p\>$ and $\sigma_{p, \alpha}$ are defined in Eq.~\ref{eq:phiPavg} and Eq.~\ref{eq:phiPvar} respectively.

\subsection{Covariance between production and depletion flux}
From the above calculation, it is easy to show that to leading order in $M$ that the covariance between the production and depletion flux is zero. To see this, we use Eq.~\ref{def:deltaPhi_p} and insert this into the definition of the covariance $\<\delta \Phi_\alpha^d  \delta \Phi_\alpha^p \> $ between the production and depletion flux. Since $\delta\Phi_\alpha^p$ has a single factor of the standard normal variable $y_{\alpha\beta}$ multiplying all terms, the only way to get a nonzero average is for $\delta\Phi_\alpha^d$ to contain factors of $y_{\alpha\beta}$ that could generate a $y_{\alpha\beta}^2$. But the depletion flux does not have any dependence on $y_{\alpha\beta}$, and so the correlation vanishes.

\section{Growth flux}

In the previous section, we focused on the production and depletion fluxes for the resources. Here, we focus on the growth flux. Because we are using the cavity method, we consider a system where a species $i$ has been removed from the ecosystem. The steady-state abundances reached by the ecosystem in the absence of species $i$ are denoted by $N_{j /i}$ and $R_{\alpha /i}$. We then ask about the growth flux that the species $i$ would have when it invades such an  ecosystem:
\be
\Phi_i^g =\sum_{\alpha} (1-l)g c_{i \alpha} R_{\alpha/i} + (1-l)g \mu_c {R_E \over M},
\ee
where once again we ignore fluctuations in the consumption of the externally supplied resource.
As before, we write $c_{i \alpha}= \mu_c/M + {\sigma_c \over \sqrt{M}} x_{i \alpha}$ to get
\begin{align}
\<\Phi_i^g\> &= (1-l)\mu_c g\left( \<R\> + {R_E \over M}\right) \nonumber \\
& \approx (1-l)\mu_c g \left(\<R\> + {R_E \over M}\right).
\label{Eq:phiGavg}
\end{align}
Defining the fluctuating part of the growth flux
\be
\delta \Phi_i^g \equiv \Phi_i^g -\<\Phi_i^g\>= \sum_\alpha {\sigma_c \over \sqrt{M}} x_{i \alpha}R_\alpha,
\ee
allows us to calculate the variance of the growth flux yielding:
\be
\sigma_{g,i}^2 \equiv \< (\delta \Phi_i^g)^2\> = (1-l)^2g^2 \sigma_c^2 \<R^2\>.
\label{Eq:phiGvar}
\ee
With these expressions, within the replica-symmetry ansatz the growth flux can be though of as a normal random variable of the form
\be
\Phi_i^g = (1-l)\mu_c g\<R\> + \sigma_{g,i} z_i^g
\ee

\section{Cavity Corrections}
The essence of the cavity calculations is to relate an ecosystem with $(S+1, M+1)$ species and resources to an ecosystem with $(S, M)$ system. We will use the language of adding an additional species and resource which by convention we will denote by $N_0$ and $R_0$. The assumption is that when $S, M \gg 1$, the addition of an additional species or resource is a small perturbation and we can relate the steady-state abundances in the presence and absence of the extra resource and species using perturbation theory.

\subsection{Setting-up the perturbation theory}

In the presence of the new resource $R_0$ and species $N_0$, Eq.~\ref{CRM_R} is modified to
as:
\be
0={d R_{\alpha} \over dt} = \k_\alpha +\delta k_\alpha - \omega_\alpha R_\alpha - \Phi_\alpha^{d} +\Phi_\alpha^p,
\ee
with
\be
\delta k_\alpha = - c_{0 \alpha} N_0R_0 + \sum_{j} l D_{\alpha 0} c_{j 0} R_0 N_j +\sum_{\beta} D_{\alpha \beta} c_{0 \beta}N_0 R_\beta + D_{\alpha 0}R_0 N_0
\ee
Similarly, Eq.~\ref{CRM_N} is modified to
\be
{d N_i \over dt} = N_i \left(\Phi^g_i -m_i - \delta m_i \right)
\ee
with
\be
\delta m_i = - (1-l)g c_{i0}R_0.
\ee

We can use perturbation theory to relate the species and resource abundances an ecosystem without the new resource and species (denoted by $N_{i /0}$ and $R_{\alpha /0}$ respectively) to the abundances in the presence of $N_0$ and $R_0$ (denoted by $N_i$ and $R_\alpha$ respectively) . To do so, we define the susceptibilities
\be
\nu_i = -{\partial N_i  \over \partial m_i}.
\ee
and
\be
\chi_\alpha  = {\partial R_\alpha \over \partial \kappa_\alpha}.
\ee
We note that this latter susceptibility differs from the resource susceptibility in \cite{cui2020effect} where we took the derivative with respect to $\omega_\alpha$
rather than $\kappa_\alpha$.

Combining the expressions above yields
\bea
R_\alpha &=& R_{\alpha /0} +  \delta R_\alpha = R_{\alpha /0} + \chi_\alpha \delta \k_\alpha \nonumber \\
 &=&R_{\alpha /0} + \chi_\alpha \left( -c_{0 \alpha} N_0R_\alpha + l \sum_{j} D_{\alpha 0} c_{j 0} N_j R_0
+ l \sum_{\beta}  D_{\alpha \beta} c_{0 \beta}R_\beta N_0 \right)
\eea
and
\bea
N_i &=& N_{i /0} + \delta N_i \nonumber= N_{i /0} -\nu_i \delta m_i \\
 &=&N_{i /0} + \nu_i  (1-l) g c_{i 0} R_0.
\eea
In writing these expressions, we have ignored the ``off-diagonal'' susceptibilities whose contributions we have assumed scale lower order in $M$.

\subsection{Computing the TAP corrections}
To proceed, we need to calculate the TAP corrections to the three fluxes $\Phi_\alpha^p$, $\Phi_\alpha^d$, and $\Phi_i^g$ due to the addition of $N_0$ and $R_0$. Let us denote these cavity corrections to the fluxes by $\Delta \Phi$. Notice through an abuse of notation we will drop the $/0$ notation so that $R_{\alpha /0}$ is represented by $R_\alpha$. For the growth flux this amounts to calculating
\bea
\Delta \Phi_0^g &=& \sum_{\alpha}(1-l_\alpha) g_{0\alpha} c_{0 \alpha} \delta R_\alpha \nonumber \\
&=& \sum_{\alpha} (1-l) g \chi_\alpha \left(- \<c_{0 \alpha} c_{0 \alpha}\> N_0R_\alpha + l \sum_{j} D_{\alpha 0} \<c_{0 \alpha} c_{j 0}\> N_j R_0
+ l \sum_{\beta}  D_{\alpha \beta}  \< c_{0 \alpha} c_{0 \beta} \> R_\beta N_0 \right) \nonumber \\
&=&(1-l) g\left[-{1 \over M} \sum_{\alpha}  \chi_\alpha \sigma_c^2 N_0R_\alpha + {l \over M}\chi_0  D_{0 0} \sigma_c^2 N_0 R_0
+ {1\over M} \sum_{\alpha} \sigma_c^2 \chi_\alpha D_{\alpha \alpha}R_\alpha N_0 \right]\nonumber \\
&=&-(1-l) g \sigma_c^2 N_0  \<\chi R\>
\eea
where 
\bea
\<\chi R\> = {1 \over M} \sum_{\alpha\neq E} \chi_\alpha R_\alpha
\eea
and we have dropped the last two terms from the third line because they are higher order in $1/M$. We also need
\bea
\Delta \Phi_0^d &=& \sum_j c_{j 0}  \delta N_j R_0 \nonumber \\
&=& R_0^2 \sum_j c_{j 0}^2 \nu_j (1-l) g_{j 0}  \nonumber \\
&=& \gamma^{-1}R_0^2 (1-l)g \sigma_c^2  \nu \nonumber \\
&\equiv& R_0^2 \tilde{\nu}
\eea
where we have introduced the notation
\bea
\nu &=& \frac{1}{S}\sum_j \nu_j\\
\tilde{\nu} &=& \gamma^{-1} (1-l)g \sigma_c^2 \nu.
\label{def:tildenu1}
\eea
Finally, 
\be
\Delta \Phi_0^p =  l \sum_{j,\beta}  D_{0 \beta} c_{j \beta}(R_\beta \delta N_j +\delta R_\beta N_j )
\ee
One can show that to leading order $M$ that these terms disappear in the thermodynamic limit so that one has
\be
\Delta \Phi_0^p \approx 0.
\ee


\section{Self-consistency equation}
\subsection{Equations for species} 

Let us start by defining an equation for species 0 at steady-state. We separate out the actual steady-state growth flux $\Phi_0^g$ for species 0 into three parts: the average growth flux $\langle \Phi^g \rangle$ over all species in the regional pool, a Gaussian random variable with variance $\sigma_N^2 = Var[\Phi_i^g]$ to account for the variability in growth rates in the regional pool, and finally the TAP correction $\Delta \Phi_0^g$ that accounts for the specific feedback from species 0 on the rest of the community:
\bea
0 &=& N_0( \<\Phi_0^g\> + \sigma_N  z_0^g + \Delta \Phi_0^g -m_0 ) \nonumber \\
0 &=& N_0\left[(1-l)\mu_c g\<R\>  + \sigma_N  z_0^g -(1-l) g \sigma_c^2 N_0  \<\chi R\>  - m_0\right]
\eea
Requiring that the steady-states cannot be invaded (i.e are ecologically stable) translates into the statement that we choose $N_0$ to be
\be
N_0 = \mathrm{max}\,  \left[0, {(1-l)\mu_c g\<R\>-m + \sigma_N  z_0^g \over (1-l) g \sigma_c^2\<\chi R\> } \right],
\label{Eq:defN0}
\ee
with
\be
\sigma_N^2 = \sigma_m^2+ \sigma_{g,0}^2=\sigma_m^2+ (1-l)^2\sigma_c^2g^2 \<R^2\>, 
\ee
where in writing this equation we have used the fact that $m_0$ is a normal random variable with variance $\delta m^2$ that is uncorrelated with $\Phi_0^g$ and
Eq.~\ref{Eq:phiGvar}.

Assuming replica symmetry, we can now write down the self-consistency equations for the moments of $N_i$, using the  special functions $w_n$ defined
as
\be
w_n(\Delta)= \int_{-\Delta}^\infty {1 \over \sqrt{2\pi} }e^{-z^2 \over 2}(z+\Delta)^n.
\ee
Using the usual arguments for replica-symmetric cavity solutions and matching the moments of Eq~\ref{Eq:defN0}, one arrives at the expressions for the fraction of surviving species $\phi_N$ and the first and second moments of the species abundance distributions
$\<N\>$ and $\<N^2\>$ \cite{advani2018statistical, bunin2017ecological, barbier2018generic}:
\begin{align}
\phi_N &= w_0(\Delta_N)\\
\langle N\rangle &= \frac{\sigma_{N}}{(1-l) g\sigma_c^2  \<\chi R\> } w_1(\Delta_N)\label{eq:N}\\
\langle N^2 \rangle &= \left( \frac{\sigma_{N}}{(1-l)g\sigma_c^2 \<\chi R\> }\right)^2 w_2(\Delta_N)
\end{align}
where
\begin{align}
\Delta_N = \frac{(1-l)\mu_c g\<R\>-m}{\sigma_{N}}.\label{eq:DelN}.
\end{align}
To calculate the average susceptibility we note that 
\begin{align}
\nu = -\left\langle \frac{\partial N_0}{\partial m_0}\right\rangle =  \frac{\phi_N}{(1-l) g \sigma_c^2  \<\chi R\>},
\end{align}
and, from Eq~\ref{def:tildenu1} that:
\begin{align}
\tilde{\nu} = \frac{\gamma^{-1} \phi_N}{\langle \chi R\rangle}.
\label{def:tildenu2}
\end{align}

Eq.~\ref{Eq:defN0} also allows us to compute the expectation values that occur in Eq.~\ref{eq:phiPavg} and Eq.~\ref{eq:phiPvar}. To do so, we note that
from 
\begin{align}
\sigma_{N} z_i^g = \delta m_i + (1-l) \frac{\sigma_c}{\sqrt{M}}\sum_\beta g_{i\beta} x_{i\beta} R_{\beta/i} 
\end{align}
Using this expression and the definition of $x_{i\beta}$, we obtain: 
\begin{align}
\<NxR\> &= \phi_N { \<R^2\> \over \sqrt{M} \sigma_c \<\chi R\>} \\
\<NxR^2\> &= \phi_N { \<R^3\> \over \sqrt{M} \sigma_c \<\chi R\>} \\
\<NxR^2xN\> &= \phi_N^2 { \gamma\<R^2\>^2+\<R^4\> \over M \sigma_c^2 \<\chi R\>^2} 
\label{Eq:correlationfunctions1}
\end{align}
The first three quantities have the correct scaling for the production flux and its variance to remain finite as $M\to\infty$. In calculating these expressions,
we have ignored correlation between $x_{i\alpha}$ and $R_\alpha$ which we have assumed are higher order in $1/M$. We check that this assumption is reasonable numerically.

With these expressions, we can rewrite Eqs.~\ref{eq:phiPavg} and ~\ref{eq:phiPvar} as
\be
\<\Phi_\alpha^p\> = \gamma^{-1} lD \left[ \mu_c \<N\>(\<R\>+ {R_E \over M})+\phi_N {\<R^2\> \over  \<\chi R\>}\right]
\label{Eq:phiPavg2}
\ee
and
\be
\sigma_{p, \alpha}^2 = \gamma^{-2} l^2 \sigma_D^2 \left[\mu_c^2 \<R^2\>\<N\>^2+  \phi_N^2 { (\gamma+1)\<R^2\>^2+\<R^4\> \over  \<\chi R\>^2}+ 2\mu_c  \<N\> \phi_N { \<R^3\> \over \<\chi R\>} \right].
\label{Eq:phiPvar2}
\ee
\subsection{Equations for resources}

Let us now think about the steady-state equation for resource 0. We will separate the fluxes out into the same three components -- the mean over the regional pool, a random term with variance equal to the variance over the regional pool, and a feedback term. Since we already showed that the covariance between the production and depletion fluxes vanishes to leading order in $1/M$, we can introduce independent random terms for each of these fluxes. We recall that $\kappa_0 =0$ since we are considering a scenario with just a single externally supplied resource, but we will keep the term present for the sake of defining the susceptibilities. This gives:
\bea
0 &=&  \kappa_0-\omega_0 R_0 -\<\Phi_{0}^d\> + \<\Phi_{0}^p\> - \tilde{\sigma}_d R_0 z_0^d + \sigma_{p,ex} z_0^p - \Delta \Phi_0^d + \Delta \Phi_0^p \nonumber \\
0 &\equiv & \kappa_0^\mathrm{eff} - \omega^\mathrm{eff}_0 R_0 - \tilde{\nu} R_0^2,
\eea
with 
\begin{align}
\kappa_0^\mathrm{eff}&= \kappa_0+\< \Phi_0^p\>+\sigma_p z_0^p= \gamma^{-1} lD \left( \mu_c \<N\>(\<R\>+ {R_E \over M})+\sqrt{M}\sigma_c \<NxR\> \right) +\sigma_p z_0^p\\
\omega^\mathrm{eff}_0 &= \omega +\<\Phi_0^d\> +\tilde{\sigma}_d z_0^d=\omega + \gamma^{-1}\mu_c \< N\> +\tilde{\sigma}_d z_0^d,
\label{Eq:defeffconstants}
\end{align}
where in first line we have used $\kappa_0=0$ and Eq.~\ref{eq:phiPavg} and in the second equation we have used Eq.~\ref{eq:phiDavg}. This is identical to the case without cross-feeding considered in \cite{cui2020effect} with $\kappa \rightarrow \kappa^\mathrm{eff}$, $\omega \to \omega^{\rm eff}$

Solving for $R_0$ gives:
\begin{align}
R_0 &= \frac{-\omega^{\rm eff}_0 + \sqrt{(\omega^{\rm eff}_0)^2 + 4\kappa^{\rm eff}_0 \tilde{\nu}}}{2\tilde{\nu}}.\label{eq:R1}
\end{align}
This expression allows us to compute self-consistency equations for the first two moments of the distribution of $R_\alpha$, which require performing some expansions in small parameters to solve, as in the case without crossfeeding. We can also compute
\begin{align}
\chi_0 &= \frac{1}{\sqrt{(\omega^{\rm eff}_0)^2 + 4\kappa^{\rm eff}_0 \tilde{\nu}}}
\end{align}
and find
\begin{align}
\< \chi R\> &= \frac{1}{2\tilde{\nu}}\left \langle 1 - \frac{\omega^{\rm eff}_\alpha}{\sqrt{(\omega^{\rm eff}_0)^2 + 4\kappa^{\rm eff}_0 \tilde{\nu}}}\right\rangle .
\label{eq:chiR}
\end{align}

\subsection{Solving the equations for resources}

Unfortunately, due to the non-linear nature of Eq.~\ref{eq:R1} it is not possible to solve for the moments of $R_0$ without making additional approximations. 
In what follows, we will perform an expansion assuming the fluctuations in the production flux $\Phi_p^0$ (Eqs.~\ref{eq:phiPavg} and ~\ref{eq:phiPvar})and depletion flux $\Phi_d^0$ (Eqs.~\ref{eq:phiDavg} and ~\ref{eq:phiDvar}) are small.
The first thing we need to do in order to solve these equations is to perform a Taylor expansion of the square root. To do so, we define three ``dimensionless'' parameters:
\begin{align}
\epsilon_p^2 &= \frac{ \<(\delta \Phi_0^p)^2\>}{ \<\Phi_0^p\>^2}=\frac{\sigma_p^2}{\langle \kappa^{\rm eff}\rangle^2}\\
\epsilon_d^2 &=\frac{ \<(\delta \Phi_0^d)^2\> }{\<\Phi_0^d\>^2} \times {\<R\>^2 \over \<R^2\>}= \frac{\tilde{\sigma_d}^2}{\langle \omega^{\rm eff}\rangle^2}\\
\epsilon_\kappa^1 &= \frac{\langle \kappa^{\rm eff}\rangle \tilde{\nu}}{\langle \omega^{\rm eff}\rangle^2}
\label{Eq:defepsilons}
\end{align}
From these definitions, it is clear that  the first two parameters measure fluctuations in the production and depletion flux respectively. The parameter $\epsilon_k$ is a little harder to interpret. We will see below that the leading order contribution to $\epsilon_\kappa^2$ is the just the species-packing fraction and is always less than $1/2$ (see derivation of species packing bound below). 

In terms of these parameters, we have:
\begin{align}
\sqrt{(\omega_\alpha^{\rm eff})^2 + 4 \kappa_\alpha^{\rm eff}\tilde{\nu}} = \langle \omega^{\rm eff}\rangle \sqrt{1 + 2 \epsilon_d z_\alpha^d + \epsilon_d^2 (z_\alpha^d)^2 + 4\epsilon_\kappa^2(1+\epsilon_p z_\alpha^p)}.
\end{align}
We will use the formula:
\begin{align}
\sqrt{1+x} = 1 + \frac{1}{2}x - \frac{1}{8}x^2 + \frac{1}{16}x^3 -\frac{5}{128}x^4 + \mathcal{O}(x^5).
\end{align}
Keeping terms up to order $\epsilon^4$ (and noting that several terms cancel which greatly simplifies the final expression), we have
\begin{align}
\sqrt{(\omega_\alpha^{\rm eff})^2 + 4 \kappa_\alpha^{\rm eff}\tilde{\nu}}  &= 1 +\epsilon_d z_\alpha^d  + 2\epsilon_\kappa^2[1+\epsilon_p z_\alpha^p - \epsilon_d z_\alpha^d+\epsilon_d^2(z_\alpha^d)^2 \\
 &- \epsilon_d  \epsilon_p z_\alpha^p z_\alpha^d - \epsilon_\kappa^2]  +\mathcal{O}(\epsilon^5).\label{eq:ex}
\end{align}

Using the expansion in eq.~(\ref{eq:ex}) in the expression for $R_\alpha$ found in eq.~(\ref{eq:R1}) above, we have:
\begin{align}
R_\alpha &= \frac{\langle \omega^{\rm eff}\rangle}{2\tilde{\nu}}\left(2\epsilon_\kappa^2 [1+\epsilon_p z_\alpha^p - \epsilon_d z_\alpha^d+\epsilon_d^2(z_\alpha^d)^2 - \epsilon_d  \epsilon_p z_\alpha^p z_\alpha^d - \epsilon_\kappa^2]  +\mathcal{O}(\epsilon^5)\right).
\end{align}
Now we can use the definition of $\epsilon_\kappa$ to simplify the expression further:
\begin{align}
R_\alpha &= \frac{\langle \kappa^{\rm eff}\rangle}{\langle \omega^{\rm eff}\rangle} [1+\epsilon_p z_\alpha^p - \epsilon_d z_\alpha^d+\epsilon_d^2(z_\alpha^d)^2 - \epsilon_d  \epsilon_p z_\alpha^p z_\alpha^d - \epsilon_\kappa^2]  +\mathcal{O}(\epsilon^3).\label{eq:R2}
\end{align}
This expression makes sense, because to lowest order the resource abundance is just the ratio of supply to depletion.

We can immediately calculate the expectation of various moments. The mean resource abundance is given by
\begin{align}
\langle R\rangle &= \frac{\langle \kappa^{\rm eff}\rangle}{\langle \omega^{\rm eff}\rangle}( 1+ \epsilon_d^2 - \epsilon_k^2) +\mathcal{O}(\epsilon^3).
\label{eq:avgR1}
\end{align}
Similarly, one can express higher order moments as
\begin{align}
\langle R^2 \rangle &=\<R\>^2( 1+ \epsilon_d^2 + \epsilon_p^2)  +\mathcal{O}(\epsilon^3)\nonumber\\
\langle R^3 \rangle &=\<R\>^3( 1+ 3\epsilon_d^2 + 3\epsilon_p^2)  +\mathcal{O}(\epsilon^3)\nonumber\\
\langle R^4 \rangle &=\<R\>^4( 1+ 6\epsilon_d^2 + 6\epsilon_p^2)  +\mathcal{O}(\epsilon^3)
\label{eq:rmoments}
\end{align}

\subsection{Average abundance of leaked resources}

We now solve for the average resource abundance $\<R\>$.
To proceed, we substitute the expression for $\tilde{\nu}$ in Eq.~\ref{def:tildenu2} into Eq.~\ref{eq:chiR} to obtain the intermediate relation
\begin{align}
\gamma^{-1}\phi_N &= \frac{1}{2}\left \langle 1 - \frac{\omega^{\rm eff}_0}{\sqrt{(\omega^{\rm eff}_0)^2 + 4\kappa^{\rm eff}_0 \tilde{\nu}}}\right\rangle\label{eq:bound}
\end{align}
Using eq.~(\ref{eq:ex}), this can be rewritten as
\begin{align}
\gamma^{-1}\phi_N &= \frac{1}{2}\left\langle 1 - \frac{1}{ 1 +\epsilon_d z_\alpha^d  + 2\epsilon_\kappa^2[1+\epsilon_p z_\alpha^p - \epsilon_d z_\alpha^d+\epsilon_d^2(z_\alpha^d)^2 - \epsilon_d  \epsilon_p z_\alpha^p z_\alpha^d - \epsilon_\kappa^2]  +\mathcal{O}(\epsilon^5)}\right\rangle\\
&= \frac{1}{2}\left\langle \epsilon_d z_\alpha^d  + 2\epsilon_\kappa^2 - \epsilon_d^2(z_\alpha^d)^2 \right\rangle+\mathcal{O}(\epsilon^3)\\
&=\epsilon_k^2-{\epsilon_d^2 \over 2}
\label{Eq:defepsilonk}
\end{align}
Explicitly this last equation can be written as
\be
\gamma^{-1}\phi_N= \frac{\langle \kappa^{\rm eff}\rangle\tilde{\nu}}{\langle \omega^{\rm eff}\rangle^2} - \frac{\tilde{\sigma_d}^2}{2\langle \omega^{\rm eff}\rangle^2}+\mathcal{O}(\epsilon^3).
\ee
Rearranging this equation gives
\be
\tilde{\nu} = \frac{\langle \omega^{\rm eff}\rangle^2 \gamma^{-1} \phi_N + \frac{1}{2} \tilde{\sigma}_d^2}{\langle \kappa^{\rm eff}\rangle}.
\ee
Using the expression for
$\<R\>$ in Eq.~\ref{eq:R2}, we can simplify this to:
\begin{align}
\tilde{\nu} \< R\> = \< \omega^{\rm eff}\>\left(\gamma^{-1}\phi_N + \frac{1}{2}\epsilon_d^2\right) +\mathcal{O}(\epsilon^3).
\end{align}
Combining this with Eq.~\ref{def:tildenu2} also yields the relation:
\begin{align}
\langle \chi R\rangle &= \frac{\gamma^{-1}\phi_N}{\tilde{\nu}} = \frac{\gamma^{-1}\phi_N\<R\>}{ \< \omega^{\rm eff}\>\left(\gamma^{-1}\phi_N + \frac{1}{2}\epsilon_d^2\right)}.\label{eq:chiR2}
\end{align}

Substituting our expression for $\tilde{\nu}\< R\>$, Eq.~\ref{Eq:correlationfunctions1}, and Eq.~\ref{eq:avgR1} into eq.~(\ref{Eq:defeffconstants}) gives
\begin{align}
\<\kappa_0^\mathrm{eff}\>&= \gamma^{-1} lD \left( \mu_c \<N\>(\<R\>+ {R_E \over M})+\sqrt{M}\sigma_c \<NxR\> \right) \nonumber \\
&= \gamma^{-1} lD \left( \mu_c \<N\>(\<R\>+ {R_E \over M})+ {\phi_N \<R^2\> \over \<\chi R\>} \right) \nonumber \\
&=\gamma^{-1} lD \left( \mu_c \<N\>(\<R\>+ {R_E \over M} \right)+  \gamma \< \omega^{\rm eff}\>\left(\gamma^{-1}\phi_N + \frac{1}{2}\epsilon_d^2\right){ \<R^2\> \over \<R\>} \nonumber \\
&=\gamma^{-1} lD \left( \mu_c \<N\>(\<R\>+ {R_E \over M} \right)+  \gamma \< \omega^{\rm eff}\>\left(\gamma^{-1}\phi_N + \frac{1}{2}\epsilon_d^2\right)\<R\>(1+
\epsilon_d^2 + \epsilon_p^2).
\label{Eq:defk0final}
\end{align}
We can substitute this equation into Eq.~\ref{eq:avgR1} to get
\begin{align}
\< R\> &= {R_E \over M}{q \over 1-q}\left( 1+ {\epsilon_d^2 \over 1-q} - {(1+ \gamma)\epsilon_k^2 \over 1-q}\right) + \mathcal{O}(\epsilon^3) \nonumber\\
q &=lD\left( 1-{\omega \over \<\omega_{eff}\>} \right)
\label{Eq:avgRsecondorder}
\end{align}
Substituting Eq.~\ref{Eq:REdef} into the expression above yields
\be
\< R\> = {R_E \over M}{q \over 1-q}= {\tilde{\kappa}_E \over \<\omega_{eff}\>}{q \over 1-q} +\mathcal{O}(\epsilon^2).
\label{Eq:FinalRavg1}
\ee

One final approximation we can make is consider the limit where depletion is dominated by microbes and we expand the expression above
dropping all terms  $\mathcal{O}\left({\omega \over \<\omega_{eff}\>}\right)$. In this case, defining 
\be
q_0=lD
\ee
and using energy conservation relation Eq.~\ref{Eq:EnergyCons} with $\mathrm{w}=1$ we have
that
\begin{align}
\< R\> &= {\tilde{\kappa}_E \over \gamma^{-1}\mu_c \<N\>}{q_0 \over 1-q_0} +\mathcal{O}(\epsilon^2) \nonumber \\
&= {\<N m\> \over  g\mu_c \<N\>}{q_0 \over 1-q_0} +\mathcal{O}(\epsilon^2) \nonumber \\
&={\<N\>\< m\> \over g\mu_c \<N\>}{q_0 \over 1-q_0} +\mathcal{O}(\epsilon^2) \nonumber \\
&={m \over g\mu_c}{q_0 \over 1-q_0}
\label{Eq:FinalRavg2}
\end{align}
where we have defined the expectation
\be
\<N m\>= {1 \over S} \sum_{i} N_i m_i
\ee
and in going from the second to third line we have used that $\<N m\>=\<N\>\<m\>$ at this order in $\epsilon$.


\subsection{Second and higher-order moments of resource abundance}

In order to calculate higher order moments of resources (see Eq.~\ref{eq:rmoments}) we need to calculate $\epsilon_d^2$ and $\epsilon_p^2$.
Substituting the expressions in Eq.~\ref{eq:phiDavg} and Eq.~\ref{eq:phiDvar} into Eq.~\ref{Eq:defepsilons} yields
\begin{align}
\epsilon_d^2 &=\frac{ \<(\delta \Phi_0^d)^2\> }{\<\Phi_0^d\>^2} \times {\<R\>^2 \over \<R^2\>}=\gamma {\sigma_c^2 \over \mu_c^2} {\<N^2\> \over \<N\>^2}.
\label{Eq:epsilond-final}
\end{align}
This expression implies that, as expected,  ${\sigma_c^2 \over \mu_c^2} \sim \epsilon^2$.

In order to calculate $\epsilon_p^2$, we start by calculating the average of the production flux. Combining Eq.~\ref{eq:phiPavg} with Eq.~\ref{eq:chiR} and
yields Eq.~\ref{Eq:defepsilonk} yields
\be
\<\Phi_\alpha^p\>= \gamma^{-1}\mu_c \<N\> lD(\<R\>+ {R_E \over M}) + \gamma \<\omega^{eff} \> {\<R^2\> \over \<R\>} \epsilon_k^2 +\mathcal{O}(\epsilon^3).
\ee
To simplify this expression further, we note that $\gamma^{-1}\mu_c \<N\>= \<\omega^{eff}\>-\omega$ and then use Eq~\ref{Eq:avgRsecondorder} 
to get 
\be
\<\Phi_\alpha^p\>=\<\omega^{eff} \>\<R\> + \gamma \<\omega^{eff} \> {\<R^2\> \over \<R\>} \epsilon_k^2 +\mathcal{O}(\epsilon^3).
\ee

In order to calculate $\sigma_p^2$ in Eq.~\ref{eq:phiPavg}, we start by simplifying the expressions in Eq.~\ref{Eq:correlationfunctions1} using Eq.~\ref{eq:chiR2}, Eq.~\ref{eq:rmoments}, and Eq.~\ref{Eq:defepsilonk}, yielding:
\begin{align}
\sqrt{M} \sigma_c \<NxR\> &= \gamma \<\omega^{eff}\> \<R\> \epsilon_k^2 + \mathcal{O}(\epsilon^3) \\
\sqrt{M} \sigma_c\<NxR^2\> &= \gamma \<\omega^{eff}\> \<R\>^2 \epsilon_k^2 + \mathcal{O}(\epsilon^3)  \\
M \sigma_c^2\<NxR^2xN\> &= \gamma^2 \<\omega^{eff}\>^2 \<R\>^2(1+\gamma) \epsilon_k^2 + \mathcal{O}(\epsilon^3). 
\label{Eq:correlationfunctions2}
\end{align}
Plugging these into Eq.~\ref{eq:phiPavg} gives
\be
\sigma_p^2 = {\sigma_D^2 \over D^2} \<\omega^{eff}\>^2 \<R\>^2 \left[q^2 + q^2 (\epsilon_d^2 + \epsilon_p^2) + (l^2D^2(\gamma+1)+2qlD\gamma)\epsilon_k^2 +\mathcal{O}(\epsilon^3) \right]
\ee
and Eq~\ref{eq:phiPvar}. This yields 
\be
\epsilon_p^2 = {\sigma_D^2 \over D^2} q^2 + \mathcal{O}(\epsilon^3)
\label{Eq:epsilonp-final}
\ee
where we have noted this expression implies that as expected ${\sigma_D^2 \over D^2} \sim \epsilon^2$.
Thus, to this order in $\epsilon_p^2$ we have
\begin{align}
\epsilon_p^2 &= {\sigma_D^2 \over D^2} q^2 
\end{align}

\section{Summary of results}
\label{sec:summary}
In this section, we summarize our results from the calculation. We start by defining the parameters
\begin{align}
q &=lD\left( 1-{\omega \over \<\omega_{eff}\>} \right) \nonumber \\
\epsilon_d^2 &= \gamma {\sigma_c^2 \over \mu_c^2} {\<N^2\> \over \<N\>^2} \nonumber \\
\epsilon_p^2 &= {\sigma_D^2 \over D^2} q^2 \nonumber\\
\epsilon_k^2 &= \gamma^{-1} \phi_N + {1 \over 2} \epsilon_d^2
\label{eq:summary1}
\end{align}
The average resource abundance is given by
\begin{align}
\< R\> &= {R_E \over M}{q \over 1-q}\left( 1+ {\epsilon_d^2 \over 1-q} - {(1+ \gamma)\epsilon_k^2 \over 1-q}\right) + \mathcal{O}(\epsilon^3) \nonumber\\
 &= {\tilde{k}_E \over \<\omega^{eff}\>}{q \over 1-q}\left( 1+ {\epsilon_d^2 \over 1-q} - {(1+ \gamma)\epsilon_k^2 \over 1-q}\right) + \mathcal{O}(\epsilon^3) \nonumber\\
 &= {m \over g\mu_c}{q \over 1-q}\left( 1+ {\epsilon_d^2 \over 1-q} - {(1+ \gamma)\epsilon_k^2 \over 1-q}\right) + \mathcal{O}(\epsilon^3). 
 \label{eq:summary2}
\end{align}
The higher order resource moments are just
\begin{align}
\langle R^2 \rangle &=\<R\>^2( 1+ \epsilon_d^2 + \epsilon_p^2)  +\mathcal{O}(\epsilon^3)\nonumber\\
\langle R^3 \rangle &=\<R\>^3( 1+ 3\epsilon_d^2 + 3\epsilon_p^2)  +\mathcal{O}(\epsilon^3)\nonumber\\
\langle R^4 \rangle &=\<R\>^4( 1+ 6\epsilon_d^2 + 6\epsilon_p^2)  +\mathcal{O}(\epsilon^3) \nonumber \\
q &=lD\left( 1-{\omega \over \<\omega_{eff}\>} \right) \nonumber \\
\epsilon_d^2 &= \gamma^{-1} {\sigma_c^2 \over \mu_c^2} {\<N^2\> \over \<N\>^2} \nonumber \\
\epsilon_p^2 &= {\sigma_D^2 \over D^2} q^2 \nonumber\\
\epsilon_k^2 &= \gamma^{-1} \phi_N + {1 \over 2} \epsilon_d^2.
\label{eq:summary3}
\end{align}
The full resource distribution is given by
\begin{align}
R_\alpha &= \frac{\<R\>}{2\gamma^{-1} \phi_N + \epsilon_d^2}\left[\sqrt{(1+\epsilon_d z_\alpha^d)^2 +2(2\gamma^{-1}\phi_N+\epsilon_d^2)(1+\epsilon_p z_\alpha^p)}-1-\epsilon_d z_\alpha^d\right],
\label{eq:summary4}
\end{align}
where are before $z_\alpha^d$ and $z_\alpha^p$ are standard random normal variables.

In addition, we have expressions for the species abundances.
\begin{align}
\phi_N &= w_0(\Delta_N) \nonumber \\
\langle N\rangle &= \frac{\sigma_{N}}{(1-l) g\sigma_c^2  \<\chi R\> } w_1(\Delta_N)\nonumber \\
\langle N^2 \rangle &= \left( \frac{\sigma_{N}}{(1-l)g\sigma_c^2 \<\chi R\> }\right)^2 w_2(\Delta_N) \nonumber \\
w_n(\Delta)&= \int_{-\Delta}^\infty {1 \over \sqrt{2\pi} }e^{-z^2 \over 2}(z+\Delta)^n \nonumber \\
\Delta_N &= \frac{(1-l)\mu_c g\<R\>-m}{\sigma_{N}} \nonumber \\
\sigma_N^2 &=\sigma_m^2+ (1-l)^2\sigma_c^2g^2 \<R\>^2(1+\epsilon_d^2 + \epsilon_p^2).
\label{eq:summary5}
\end{align}

Finally, we have
\begin{align}
\tilde{\nu} \< R\> &= \< \omega^{\rm eff}\>\left(\gamma^{-1}\phi_N + \frac{1}{2}\epsilon_d^2\right) +\mathcal{O}(\epsilon^3) \nonumber \\
\langle \chi R\rangle &= \frac{\gamma^{-1}\phi_N}{\tilde{\nu}} = \frac{\gamma^{-1}\phi_N\<R\>}{ \< \omega^{\rm eff}\>\left(\gamma^{-1}\phi_N + \frac{1}{2}\epsilon_d^2\right)}.
\end{align}

\section{Species Packing Bounds}

The expressions above allow us to define a simple bound for the fraction of metabolic niches that an ecosystem can occupy \cite{cui2020effect}. If we denote the number of surviving species by $S^*$ and the number of metabolites by $M$, then in the thermodynamic limit the species-packing fraction is just $S^*/M=
\gamma^{-1} \phi_N$. To derive the bound, we start with Eq.~\ref{eq:chiR} which reads
\begin{align}
\< \chi R\> &= \frac{1}{2\tilde{\nu}}\left \langle 1 - \frac{\omega^{\rm eff}_0}{\sqrt{(\omega^{\rm eff}_0)^2 + 4\kappa^{\rm eff}_0 \tilde{\nu}}}\right\rangle .
\end{align}
Substituting Eq.~\ref{def:tildenu2} for $\tilde{\nu}$ into the expression yields
\begin{align}
\gamma^{-1}\phi_N &= \frac{1}{2}\left \langle 1 - \frac{\omega^{\rm eff}_0}{\sqrt{(\omega^{\rm eff}_0)^2 + 4\kappa^{\rm eff}_0 \tilde{\nu}}}\right\rangle\label{eq:bound}.
\end{align}
Since the second term is always positive we have that 
\be
\gamma^{-1}\phi_N \le \frac{1}{2},
\ee
proving our bound. Note that in proving this bound, we have made no approximations beyond replica-symmetry and therefore expect this to be exact in the thermodynamic limit.

\section{Fraction of surviving species depends only on competition and not leakage rate}

In this section, we show that to leading order in $\epsilon$ (see Eq.~\ref{Eq:defepsilons}), the fraction of surviving species depends only on the competition between species  ($\mu_c$ and $\sigma_c$) but not the leakage rate $l$ when $\sigma_m^2$, the fluctuations in the maintenance costs, can be ignored. Technically, this is a statement that  $\Delta_N$, and hence $\phi_N$, depends on the variance over the mean squared of consumer preferences $\sigma_c^2/\mu_c^2$ but not the leakage parameter $l$. There are many ways to show this and here we present one particularly simple derivation of this fact.

Our starting point is Eq.~\ref{eq:summary5}: 
\be
\langle N\rangle = \frac{\sigma_{N}}{(1-l) g\sigma_c^2  \<\chi R\> } w_1(\Delta_N).
\ee
We then use Eq.~{eq:chiR2} to rewrite this as
\begin{align}
\langle N\rangle &= {\sigma_{N} \over (1-l) g\sigma_c^2  \<\chi R\> } \nonumber\\
&={\sigma_{N} \< \omega^{\rm eff}\>\epsilon_k^2 \over (1-l) g\sigma_c^2 \gamma^{-1}\phi_N\<R\>},
\end{align}
where in the second line we have used Eq.~\ref{eq:chiR2} and Eq.~\ref{eq:summary1} for $\epsilon_k^2$. Now, using the fact that when $\sigma^2=0$ we have from Eq.~\ref{eq:summary5} that
\be
\sigma_N= (1-l)\sigma_c g \<R\> (1+{1 \over 2}\epsilon_d^2 +{1 \over 2} \epsilon_p^2) + \mathcal{O}(\epsilon^3),
\ee
the above can be rewritten as
\begin{align}
\langle N\rangle &={ \< \omega^{\rm eff}\> \epsilon_k^2 \over \sigma_c \gamma^{-1}\phi_N }+ \mathcal{O}(\epsilon^3) \nonumber \\
\<N\>&=\gamma^{-1}\<N\>{ \mu_c \over \sigma_c} \left(1+  {\epsilon_d^2 \over 2 \gamma^{-1}\phi_N }\right)+ \mathcal{O}(\epsilon^3) \nonumber \\
\<N\>&=\gamma^{-1}\<N\>{ \mu_c \over \sigma_c} \left(1+ {\sigma_c^2 \over \mu^2} {\<N^2\> \over 2 \phi_N \<N\>^2 }\right)+ \mathcal{O}(\epsilon^3) \nonumber \\
\<N\>&=\gamma^{-1}\<N\>{ \mu_c \over \sigma_c} \left(1+ {\sigma_c^2 \over \mu^2} {w_2(\Delta_N) \over 2 w_0(\Delta_N) w_1(\Delta_N)^2 }\right)+ \mathcal{O}(\epsilon^3)
\label{Eq:cavitynumeric1}
\end{align}
Canceling the $\<N\>$ from both sides yields 
\be
1 =\gamma^{-1}{ \mu_c \over \sigma_c} \left(1+ {\sigma_c^2 \over \mu^2} {w_2(\Delta_N) \over 2 w_0(\Delta_N) w_1(\Delta_N)^2 }\right)+ \mathcal{O}(\epsilon^3),
\ee
showing that $\Delta_N$ and hence $\phi_N$ only depends on ${\sigma_c^2 \over \mu_c}$ at this order.

\section{Details of numerical simulations}

\subsection{Numerical Simulations}
We now give an overview of how we performed numerical simulations. All code can be found on the corresponding Github page. To numerically simulate the communities, we used the Community Simulator package to find steady-states with indicated parameters \cite{ marsland2020community}. We used the parameters: $q=0$ (generalist consumers), supply = 'external', regulation= 'independent', response='type I' , and sparsity $s=0.01$. These choices ensure that the dynamics follow Eq.~\ref{CRM_R}. We set the number of species $S=600$, the number of metabolites equal $M=300$, the average maintenance cost $m=1$ with standard deviation $\sigma_m=0.1$. A single resource was
supplied externally at a rate $\tilde{\kappa}_E =\kappa_E/M=100$ and $\omega=1$. The leakage rate was chosen to be $l=0.8$ if fixed, or varied over $10$ equal intervals from $l=0.7-0.95$ as indicated in figures. 

The consumption coefficients were chose from binary
distribution with $c_{i \alpha} \in \{0, c_1\}$ with probability having non-zero coefficient $c_1$ given by $p=\mu_c/(c_1M)$. This was done to ensure that the average consumption rate $\<c_{i \alpha}\>= \mu_c/M$, with $\mu_c=1$. Under these assumptions $\sigma_c= \sqrt{Mc_1 p(1-p)}$. In order to vary competition, $c_1$ was varied over $10$ equal interval from $c_1=1/M-10/M$ with $M=300$. Default choice for $c_1= 9/M=0.03$. See section ``Sampling parameters'' in \cite{marsland2020community} for details and Github for code.

  Averages $\phi_N$, $\<N\>$, $\<N^2\>$, $\<R\>$, and $\<R^2\>$ reflect averages over 28 independent realizations. In order to ensure numerical stability, we treat species as abundances below $10^{-3}$ as going extinct.

\subsection{Numerical solution of cavity equations}

To numerically check the solutions to our self-consistent cavity equations, we used the expressions summarized in section \ref{sec:summary}. In particular, given the parameters we solved for $\Delta_N$ numerically. To do so, we  used the second line of Eq.~\ref{Eq:cavitynumeric1},  Eq.~\ref{eq:summary3}, and Eq.~\ref{eq:summary5} to get:
\begin{align}
\phi_N{ \sigma_c \over \mu_c}&= \left( \gamma^{-1}\phi_N+  {\epsilon_d^2 \over 2 }\right) \nonumber\\
\phi_N{ \sigma_c \over \mu_c}&=\left( \gamma^{-1}\phi_N+  {\sigma_c^2 \over \mu_c^2} {\<N\>^2 \over \<N\>^2} \right)\nonumber \\
w_0(\Delta_N){ \sigma_c \over \mu_c}&=\left( \gamma^{-1}w_0(\Delta_N)+  {\sigma_c^2 \over \mu_c^2} {w_2(\Delta_N) \over w_1(\Delta_N)^2} \right)
\end{align}
This allowed us define the cost function 
\begin{align}
\mathcal{C}(\Delta_N)=\left(w_0(\Delta_N){ \sigma_c \over \mu_c}-\left( \gamma^{-1}w_0(\Delta_N)+  {\sigma_c^2 \over \mu_c^2} {w_2(\Delta_N) \over w_1(\Delta_N)^2} \right)\right)^2,
\end{align}
which we can minimize to solve for $\Delta_N$ using the ``minimize scalar'' function from the scipy.opitmize package.

Once we solve for $\Delta_N$, we can use Eq.~\ref{eq:summary1} along with Eq.~\ref{eq:summary4} to numerically calculate the 
predicted resource distribution:
\begin{align}
R_\alpha &= \frac{\<R\>}{2\gamma^{-1} \phi_N + \epsilon_d^2}\left[\sqrt{(1+\epsilon_d z_\alpha^d)^2 +2(2\gamma^{-1}\phi_N+\epsilon_d^2)(1+\epsilon_p z_\alpha^p)}-1-\epsilon_d z_\alpha^d\right].
\end{align}
To do so, we randomly sample from this distribution $10^6$ times and then plot the resulting pdf.

We can also use Eq.~\ref{Eq:defN0} to generate the distribution of the species abundances. In particular, this equation states that the distribution of 
\emph{surviving species} follows a truncated Gaussian distribution with a fraction $\phi_N$ of the species surviving. To proceed, note that from
Eq.~\ref{eq:summary5} that we can define the quantity
\be
\Sigma= {\<N\> \over w_1(\Delta_N)}= \frac{\sigma_{N}}{(1-l) g\sigma_c^2  \<\chi R\> }
\ee
In terms of $\Sigma$, we can rearrange Eq.~\ref{Eq:defN0} to derive that \emph{surviving species} will be described by the distribution
\be
P_N^{sur}(N)= {1 \over \sqrt{2\pi} \phi_N \Sigma}e^{-0.5({N \over \Sigma}-\Delta_N)^2}.
\ee
$\Delta_N$ is obtained by minimizing $\mathcal{C}(\Delta_N)$ as discussed above. To obtain $\<N\>$ to leading order in $\epsilon$, we use energy conservation Eq.~{Eq:EnergyCons}:
\begin{align}
\kappa_E  &\approx  \sum_i \frac{N_i m_i}{\mathrm{w} g} \nonumber \\
{\kappa_E  \over S}&\approx {1 \over S} \sum_i \frac{N_i m_i}{\mathrm{w} g} \nonumber \\
{\kappa_E  \over S}&\approx  \frac{\<N m\>}{\mathrm{w} g} \nonumber \\
{\kappa_E  \over S} &\approx  \frac{\<N\>\<m\>}{\mathrm{w} g}, 
\end{align}
which can be rearrange to yield 
\be
\<N\>\approx {\mathrm{w} g \kappa_E  \over m S},
\ee
allowing us to calculate $\Sigma$. 

As a further check the prediction of the cavity method and our approximations we use the value of $\Delta_N$ obtained from the procedure above
along with Eq.~\ref{eq:summary5} to compute
\be
{\<N^2\> \over \<N\>^2}={w_2(\Delta_N) \over w_1(\Delta_N)}.
\ee
To check that the average resource abundance $\<R\>$ only depend on the leakage rate  
in the limit of fast dilution (i.e. when $q \approx q_0=l$),  we plotted the leading order contribution to  Eq.~\ref{eq:summary2}:
\be
\<R\> \approx {m \over g\mu_c}{l \over 1-l}
\ee
versus the average obtained from direct numerical simulation.

\section{Additional Figures and Numerical Checks}
\begin{figure}[t]
\centering
\includegraphics[width=0.95\columnwidth]{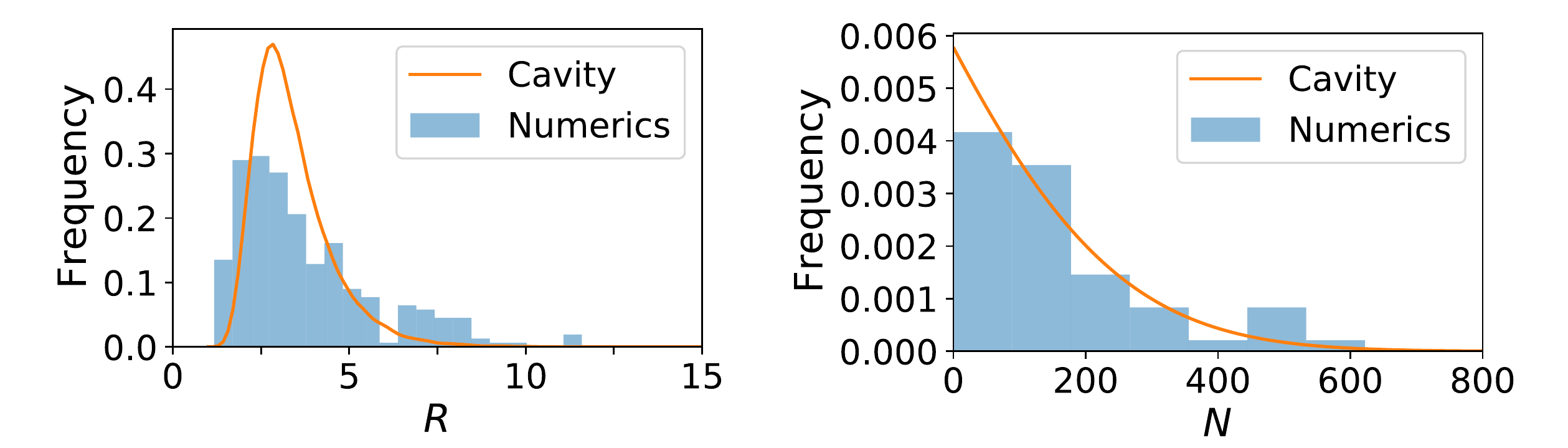}
\caption{{\bf Cavity solutions predict distributions of species and abundances.}   
The resource and species abundance distributions from numerics 
and our cavity solutions Eqs.~\ref{eq:Rdist_main} and ~\ref{Eq:defN0_main}. See appendix for parameters.}
\label{fig:distributions}
\end{figure}

\begin{figure}[h!]
\centering
\includegraphics[width=0.9\columnwidth]{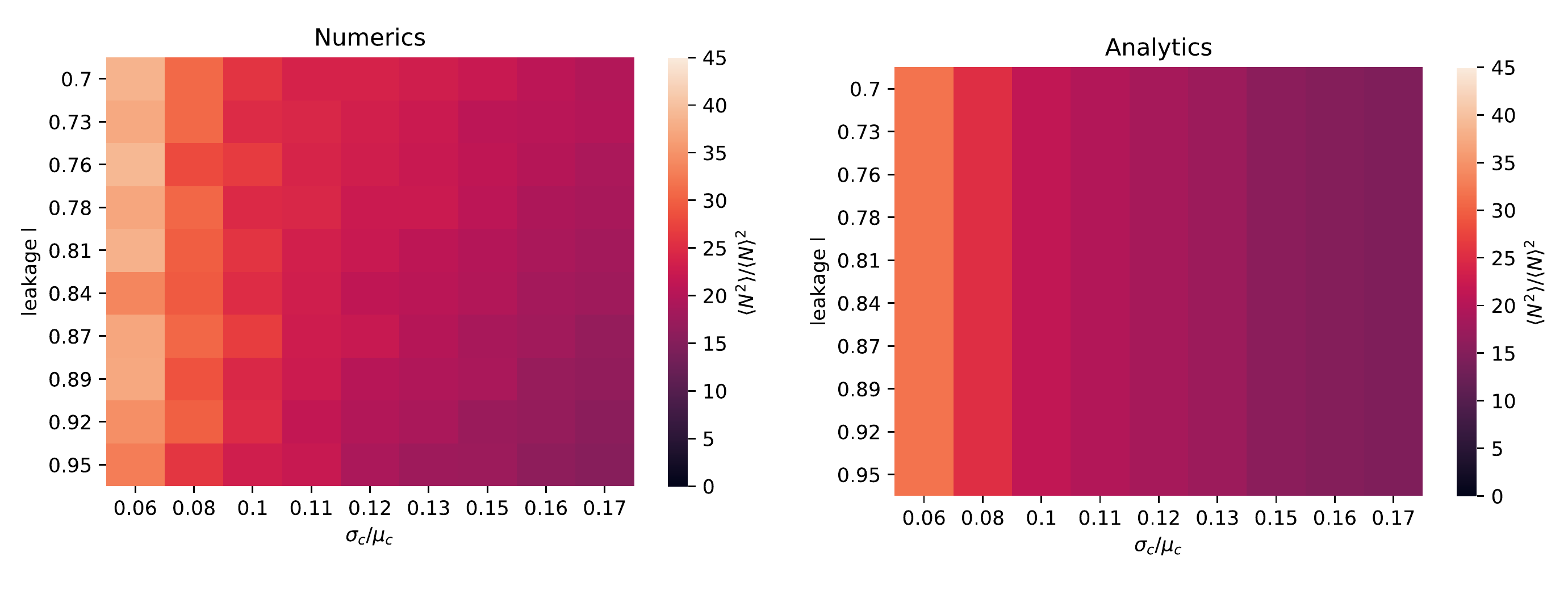}
\caption{{\bf Species distributions set by competition.} Numerical
simulations (left) and analytics (right) of $\<N^2\>/\<N\>^2$  as a function of leakage
rate $l$ and $\sigma_c/\mu_c$ given by Eq. \ref{Eq:cavitynumeric1} for binary consumer resource coefficients as described above. 
}
\label{fig:analytics_appendix}
\end{figure}

\begin{figure}[h!]
\centering
\includegraphics[width=0.5\columnwidth]{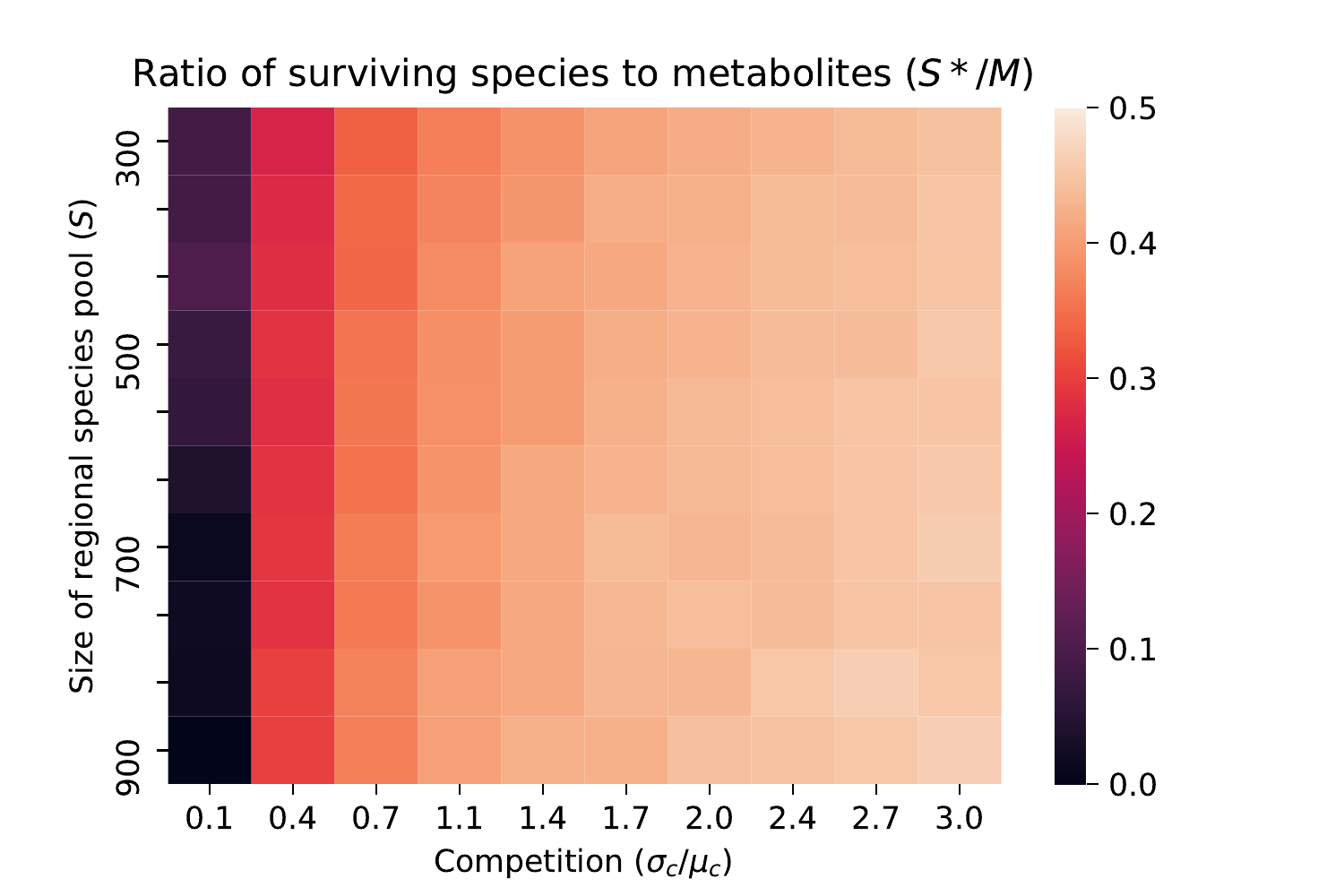}
\caption{{\bf Species packing bound with Gaussian consumer preferences.} Ratio of surviving species 
to resources (including metabolic biproducts) ${S^*/M}$ for different leakage rates $l$ and different choices of $\sigma_c/\mu_c$ with size of regional species pool $S=600$ and $M=300$. Notice that $S^*/M \le 1/2$, consistent analytic bound derived in appendix. }
\label{fig:bound_appendix}
\end{figure}

\end{document}